\documentclass[a4paper,fleqn]{cas-dc}

\usepackage[authoryear]{natbib}
\usepackage{algorithm}
\usepackage{algorithmic}
\usepackage{dsfont}
\def\tsc#1{\csdef{#1}{\textsc{\lowercase{#1}}\xspace}}
\tsc{WGM}
\tsc{QE}
\begin{document}
\let\WriteBookmarks\relax
\def\floatpagepagefraction{1}
\def\textpagefraction{.001}

\shorttitle{Active Source-Free Domain Adaptation for Medical Vision Foundation Models}    

\shortauthors{Jin Yang, Daniel S. Marcus and Aristeidis Sotiras}  

\title [mode = title]{Adapting Medical Vision Foundation Models for Volumetric Medical Image Segmentation via Active Learning and Selective Semi-supervised Fine-tuning}




\author[1]{Jin Yang}

\author[1]{Daniel S. Marcus}

\author[1,2,3,4]{Aristeidis Sotiras}[orcid=0000-0003-0795-8820]


\ead{aristeidis.sotiras@pennmedicine.upenn.edu}
\cortext[2]{Corresponding author}
\cormark[2]


\affiliation[1]{organization={Mallinckrodt Institute of Radiology},
            addressline={Washington University School of Medicine in St. Louis}, 
            city={St. Louis},
            postcode={63110}, 
            state={MO},
            country={USA}}
            
\affiliation[2]{organization={Institute for Informatics, Data Science and Biostatistics},
            addressline={Washington University School of Medicine in St. Louis}, 
            city={St. Louis},
            postcode={63110}, 
            state={MO},
            country={USA}}
            
\affiliation[3]{organization={Department of Radiology, Perelman School of Medicine},
            addressline={University of Pennsylvania}, 
            city={Philadelphia},
            postcode={19104}, 
            state={PA},
            country={USA}}

\affiliation[4]{organization={Center for AI and Data Science for Integrated Diagnostics (AI2D)},
            addressline={Perelman School of Medicine, University of Pennsylvania}, 
            city={Philadelphia},
            postcode={19104}, 
            state={PA},
            country={USA}}


\begin{abstract}
Medical vision foundation models (Med-VFMs) have recently shown strong potential for medical image analysis by leveraging representations learned from large-scale self-supervised pre-training on unlabeled medical images. However, their performance and generalization remain limited in downstream tasks, particularly volumetric medical image segmentation. While fine-tuning on labeled target-domain data improves performance, existing approaches typically rely on randomly selected samples, which may fail to identify the most informative data and thus hinder adaptation. To address these limitations, we propose an Active Selective Semi-supervised Fine-tuning (ASSFT) framework for efficient adaptation of Med-VFMs to generalize across volumetric medical image segmentation. ASSFT integrates a novel active learning strategy with selective semi-supervised learning to maximize adaptation performance under a limited annotation budget, without requiring access to source data. Specifically, we introduce an Active Test-Time Sample Query strategy that identifies informative samples from the target domain using two complementary query metrics: Diversified Knowledge Divergence (DKD) and Anatomical Segmentation Difficulty (ASD). DKD quantifies both the knowledge gap between pre-training and target domains and the semantic diversity within the target dataset, enabling the selection of samples that contain previously unlearned knowledge while maintaining intra-domain diversity. ASD estimates the segmentation difficulty of target anatomical structures by measuring predictive uncertainty within foreground regions of interest, allowing the model to prioritize samples with complex anatomical patterns rather than those dominated by background uncertainty. Second, we propose a Selective Semi-supervised Fine-tuning strategy to further improve adaptation performance by leveraging unlabeled target samples. Instead of utilizing all pseudo-labeled data, the proposed method selectively incorporates reliable unlabeled samples based on predictive confidence and semantic distance to labeled samples, enabling stable semi-supervised training while avoiding noisy pseudo-labels. We evaluated ASSFT on five volumetric medical image segmentation tasks involving different imaging modalities, anatomical structures, and dataset scales. Experimental results showed that the proposed method consistently outperformed state-of-the-art active learning and active domain adaptation approaches while requiring substantially fewer annotated samples, thus demonstrating its effectiveness, efficiency, and strong generalization capability.

\end{abstract}


%\nocite{*}

% Keywords

\maketitle

\section{Introduction}
Medical visual foundation models (Med-VFMs) have recently been developed for interpreting medical imaging and have demonstrated impressive ability across various medical image analysis tasks \citep{zhang2024challenges}. These models typically employ convolutional neural networks or vision transformers as backbone architectures and are pre-trained on large-scale unlabeled medical imaging datasets via self-supervised learning \citep{moor2023foundation,zhou2023foundation,pai2025vision,zhu20263d,wang2026vision}. After pre-training, Med-VFMs can either be directly applied to inference tasks without adaptation (i.e., non-adaptive tasks) or serve as general-purpose feature extractors that are further adapted to downstream tasks (i.e., adaptive tasks). For non-adaptive tasks such as zero-shot inference, Med-VFMs can achieve remarkable performance by leveraging the intrinsic knowledge learned during per-training. However, their performance and generalization remain limited when applied to adaptive downstream tasks, particularly medical image segmentation. In such cases, a task-specific decoder is typically incorporated with the pre-trained encoder to form an encoder–decoder segmentation network. Since the decoder is newly initialized and the network has not yet been fine-tuned using labeled samples from the target domain, the model is unable to effectively capture domain-specific anatomical patterns, resulting in inferior segmentation performance.

To improve the performance and generalization of Med-VFMs on downstream segmentation tasks, a common approach is to fine-tune the model using a subset of labeled samples from target evaluation domains. Existing studies only rely on random sampling to select these fine-tuning samples \citep{pai2025vision}. However, random selection may not be efficient because it does not guarantee that the most informative samples are used for adapting models to achieve the optimal performance. Active learning (AL) provides a feasible solution by iteratively selecting the most informative samples for annotation, thereby maximizing model performance with minimal labeling cost \citep{settles2011theories}. Existing AL methods design various query strategies to evaluate sample informativeness, often based on predictive uncertainty or sample diversity \citep{siddiqui2020viewal,wu2021redal,xie2022towards,li2023heterogeneous,yang2024plug}. Similarly, active domain adaptation (ADA) methods employ AL strategies to select informative target-domain samples for improving cross-domain generalization \citep{ning2021multi,du2023diffusion,zhang2024revisiting}.

Despite their potential, directly applying AL or ADA methods methods to adapt Med-VFMs for downstream segmentation tasks presents several limitations. First, these methods lack mechanisms to exploit the rich knowledge embedded in pre-trained Med-VFMs when evaluating sample informativeness. Leveraging this prior knowledge could enable the identification of samples that contain previously unlearned patterns in the target domain, thereby improving adaptation efficiency and generalization. Second, most existing query strategies measure informativeness at the image or volume level. In medical segmentation tasks, foreground anatomical structures typically occupy only a small portion of the image, while background regions dominate the volume. Thus, image-level informativeness estimation may be biased toward background uncertainty and underestimate informative foreground regions, limiting the ability of the model to learn important features of anatomical structures of interest \citep{wang2024patch}. Third, many ADA methods rely on access to source-domain data to during adaptation \citep{mahapatra2024alfredo}. However, the training data used for pre-training Med-VFMs are typically unavailable due to privacy constraints, making these methods difficult to apply in practice. Although active source-free domain adaptation (ASFDA) methods adapt models without the access to source data \citep{wang2023mhpl,li2023source}, their effectiveness is still constrained by the first two issues.

Another limitation of existing approaches is that they primarily rely on labeled samples selected by AL while ignoring the large number of unlabeled samples available in the target domain. Semi-supervised learning techniques can utilize these unlabeled samples by generating pseudo-labels using the current model and incorporating them with labeled samples into training \citep{wang2024dual,wang2022unsupervised,he2024enhancing}. However, using pseudo-labels of all unlabeled samples for semi-supervised training introduces two challenges. First, during early AL iterations, the model has not yet sufficiently adapted to the target domain, leading to inaccurate pseudo-labels for many samples. Training on these low-quality pseudo-labels may degrade model performance. Second, the number of unlabeled samples is often significantly larger than the number of queried labeled samples, causing pseudo-labeled data to dominate the training objective and potentially hindering the model from learning reliable patterns from labeled samples.

Overall, the central challenge in adapting Med-VFMs for segmentation is to identify samples that are simultaneously informative with respect to the model’s pre-trained knowledge and relevant to downstream anatomical structures. Existing AL strategies typically focus on uncertainty or diversity alone, which is insufficient in the context of foundation models, where prior knowledge and domain shift must both be explicitly considered. This motivates a selection strategy that jointly captures knowledge novelty, data diversity, and segmentation difficulty.

To tackle these limitations, we propose a novel \textbf{Active Selective Semi-supervised Fine-tuning (ASSFT)} framework for adapting Med-VFMs with high efficiency and generalization to target domains in volumetric medical image segmentation. The proposed method integrates a novel active learning strategy with a selective semi-supervised fine-tuning mechanism to efficiently exploit both labeled and unlabeled target-domain data. 

First, sample selection is guided by three complementary aspects: (1) knowledge novelty with respect to the pre-trained model, (2) diversity among candidate samples, and (3) segmentation difficulty related to anatomical structures. To operationalize this, we introduce an \textbf{Active Test-Time Sample Query} strategy that selects the most informative samples for annotation without requiring access to source-domain data. This strategy evaluates sample informativeness using two complementary query metrics jointly: \textbf{Diversified Knowledge Divergence (DKD)} and \textbf{Anatomical Segmentation Difficulty (ASD)}. DKD identifies samples that contain previously unlearned target knowledge relative to the pre-trained model while simultaneously enhancing semantic diversity within the target domain. In particular, DKD measures knowledge divergence between the pre-training and target adaptive distributions by identifying source-dissimilar samples, while also measuring intra-domain diversity to avoid selecting samples with overlapping patterns for fine-tuning. ASD estimates the difficulty of segmenting anatomical structures by measuring predictive entropy within foreground regions, allowing the method to prioritize samples with complex anatomical patterns that are challenging for the adaptation and generalization of the model. 

Second, we propose a \textbf{Selective Semi-supervised Fine-tuning} strategy to further improve adaptation performance and efficiency by leveraging labeled and reliable unlabeled samples. Instead of using pseudo-labels for all unlabeled data during adaptive fine-tuning, our strategy selectively identifies samples that are likely to produce high-quality pseudo-labels. Specifically, unlabeled samples are selected for semi-supervised training based on high predictive confidence and small semantic distance to labeled samples, indicating that their semantic patterns have already been partially learned by the model. By incorporating only these reliable pseudo-labeled samples together with actively queried labeled samples, the proposed method achieves more stable and effective semi-supervised fine-tuning, ultimately enhancing adaptation efficiency and generalization.

The proposed ASSFT framework is designed to be broadly applicable to different medical image segmentation scenarios (e.g, different target organs and different imaging modalities) and does not require access to source-domain data. Thus, ASSFT enables efficient and high-performing deployment of Med-VFMs in clinical settings. To evaluate its effectiveness and generalization ability, we conducted extensive experiments across five volumetric medical image segmentation tasks with different imaging modalities, anatomical structures, and dataset scales. Experimental results demonstrate that ASSFT consistently outperforms state-of-the-art AL and ADA methods while requiring significantly fewer annotated samples. Additionally, ASSFT improves the performance of Med-VFMs significantly over zero-shot segmentation on target domains even though they are pre-trained on different imaging modalities. The main contributions of this work are summarized as follows:

\begin{itemize}
    \item We propose an \textbf{Active Selective Semi-supervised Fine-tuning} framework for adaptation of medical vision foundation models in volumetric medical image segmentation with high efficiency and generalization, integrating active learning and semi-supervised learning into a unified adaptation framework.
    \item We introduce an \textbf{Active Test-Time Sample Query} strategy that selects informative samples using two novel metrics, \textbf{Diversified Knowledge Divergence} and \textbf{Anatomical Segmentation Difficulty}, enabling effective sample querying without requiring access to source-domain data.
    \item We develop a \textbf{Selective Semi-supervised Fine-tuning} strategy that improves adaptation performance and efficiency by selectively incorporating reliable pseudo-labeled samples.
    \item We conduct extensive experiments on five diverse volumetric medical image segmentation tasks, demonstrating the effectiveness, efficiency, and strong generalization capability of the proposed ASSFT framework.
\end{itemize}

\section{Related Works}
\subsection{Active Learning in Medical Image Segmentation}
Active learning has been widely explored to improve the efficiency of training medical image segmentation models by identifying the most informative samples while minimizing annotation cost. Early work, such as suggestive annotation, combines uncertainty and similarity to select representative and uncertain samples for labeling \citep{yang2017suggestive}. Bayesian approaches, including Bayesian UNet, leverage Monte Carlo dropout to estimate predictive uncertainty and prioritize samples with high uncertainty \citep{hiasa2019automated}. Similarly, query-by-committee methods estimate uncertainty using ensembles of models \citep{nath2020diminishing}. 

Beyond uncertainty-based strategies, several approaches incorporate additional criteria. For example, DSAL integrates active learning with semi-supervised learning by selecting both high- and low-uncertainty samples for strong and weak labeling, respectively \citep{zhao2021dsal}. Hybrid frameworks have been proposed to jointly consider pixel-wise entropy, regional consistency, and image diversity \citep{li2023hal}, while stochastic batch selection strategies evaluate batch-level uncertainty to improve sampling efficiency \citep{gaillochet2023active}. More recent methods focus on domain-specific criteria, such as anatomical consistency and boundary uncertainty \citep{zhou2024sbc}, image similarity for representative sampling \citep{shu2025active}, and cross-modal selection strategies for multi-domain settings \citep{chen2025active}.

Despite these advances, most existing active learning strategies focus primarily on uncertainty or diversity, or combinations thereof. However, these criteria are insufficient in the context of foundation models, where pre-trained representations encode prior knowledge that may not align with the target domain. Therefore, effective sample selection must explicitly account for both knowledge adaptation and domain shift. This motivates a selection strategy that jointly considers knowledge novelty, data diversity, and segmentation difficulty, forming the basis of our proposed approach.

\subsection{Active Source-Free Domain Adaptation in Medical Image Analysis}
Active source-free domain adaptation  has recently been explored to adapt models to target domains without access to source-domain data, particularly in medical image classification and segmentation. Early approaches combine active learning with domain adaptation to identify informative samples for annotation. For example, feature disentanglement-based methods select representative samples by separating domain-specific and task-specific features \citep{mahapatra2024alfredo}. Similarly, the Source-domain and Target-domain Dual-Reference (STDR) strategy selects samples by leveraging reference distributions from both domains to capture domain-invariant and domain-specific characteristics \citep{wang2024dual}. 

Many ASFDA methods rely on uncertainty and diversity to guide sample selection. For instance, uncertainty-guided approaches estimate predictive uncertainty to identify informative samples, often combining uncertainty with clustering-based diversity strategies \citep{luo2024uncertainty}. In vessel segmentation tasks, cascade selection strategies have been proposed to prioritize patches with high foreground certainty, focusing annotation efforts on regions of interest \citep{wang2024advancing}. More general frameworks integrate attention mechanisms and active learning strategies to support cross-task adaptation, including classification, segmentation, and detection \citep{kothandaraman2023salad}, while others combine image-level uncertainty with organ-level diversity for multi-domain and multi-modality adaptation \citep{yang2025active}.

Despite these advances, most existing ASFDA methods still rely primarily on uncertainty and diversity for sample selection. However, in the absence of source-domain data, adapting pre-trained models requires explicitly accounting for how new samples interact with the model’s existing representations. In particular, effective adaptation must consider not only sample uncertainty and diversity, but also knowledge novelty relative to the pre-trained model and task-specific factors such as segmentation difficulty. This highlights the need for a unified selection strategy that jointly captures these aspects, motivating the proposed framework.

\subsection{Medical Vision Foundation Models}
Recent years have seen a rapid growth in the development of Med-VFMs, driven by the increasing availability of large-scale medical imaging datasets and advances in self-supervised learning. These models aim to learn generalizable representations that can be transferred across tasks, modalities, and clinical applications. Representative examples include generalist medical AI models trained on diverse multi-modal datasets \citep{moor2023foundation}, as well as modality-specific foundation models such as RETFound for retinal imaging \citep{zhou2023foundation}. In volumetric imaging, several large-scale 3D foundation models have been proposed. CT-FM leverages label-agnostic contrastive learning on large collections of CT images \citep{pai2025vision} while FM-CT is pre-trained on head CT data using self-supervised learning for generalizable disease diagnosis \citep{zhu20263d}. Additionally, various foundation models are developed for 3D MR image and brain MR image analysis \citep{sun2025foundation,deng2025brain,wang2026vision,tak2026generalizable}. Therefore, these examples illustrate the diversity of architectures and pre-training strategies being explored in this rapidly evolving field. 

Despite their strong representational capabilities, adapting Med-VFMs to downstream segmentation tasks remains challenging, particularly under domain shift and limited annotation settings. Importantly, the proposed framework does not rely on specific architectural assumptions and is designed to be broadly applicable across different types of Med-VFMs. In this work, we demonstrate its effectiveness using a representative foundation model, but the approach is general and can be extended to other models and modalities.

\begin{algorithm}[tb]
\caption{Active Selective Semi-supervised Fine-tuning (ASSFT) of Med-VFMs}
\label{alg:algorithm}
\textbf{Input}: Pre-trained encoder $E^{(p)}$; initialized decoder $D^{(0)}$; unlabeled target set $\mathbb{T}=\{\boldsymbol{X}_t\}=\{x_1,.,x_t,.,x_{\mathcal{N}_t}\}$; maximum AL rounds $\mathcal{R}$; query budget $\mathcal{N}_B$; unlabeled selection budget $\mathcal{N}_{SU}$. \\
\textbf{Output}: Target-adapted segmentation network $\mathscr{F}(\boldsymbol{\Theta}^*)$. \\
\textbf{Begin}:
\begin{algorithmic}[1]
%\STATE Let current round $r=0$;
\STATE Construct a segmentation network $\mathscr{F}(\boldsymbol{\Theta})$ via $\mathscr{F}(\boldsymbol{\Theta})\leftarrow\mathscr{F}^{(0)}(\boldsymbol{\Theta};[E^{(p)};D^{(0)}])$
\STATE Initialize labeled and unlabeled sets: $\mathbb{T}_l\leftarrow\varnothing$, $\mathbb{T}_u\leftarrow\mathbb{T}$
\STATE Select $x_1$ and annotate $y_1$ to bootstrap training
\STATE Update $\mathbb{T}_l\leftarrow\mathbb{T}_l\cup\{(x_1,y_1)\}$, $\mathbb{T}_u\leftarrow\mathbb{T}_u\setminus\{x_1\}$

\FOR{$r=1$ to $\mathcal{R}$}
\STATE Compute $\textrm{DKD}(x_u)$ for $x_u\in\mathbb{T}_u$ by Eq.~3
\STATE Compute $\textrm{ASD}(x_u)$ for $x_u\in\mathbb{T}_u$ by Eq.~10
\STATE Calculate query score $Q(x_u)$ by Eq.~15
\STATE Query $\mathcal{N}_B$ samples $\boldsymbol{X}_t^r$ with highest scores for annotation $\boldsymbol{Y}_t^r$
\STATE Update sets: $\mathbb{T}_l\leftarrow\mathbb{T}_l\cup\{(\boldsymbol{X}_t^r,\boldsymbol{Y}_t^r)\}$, $\mathbb{T}_u\leftarrow\mathbb{T}_u\setminus\{\boldsymbol{X}_t^r\}$
\STATE Fine-tune the network $\mathscr{F}^{(r)}(\boldsymbol{\Theta})$ using labeled set $\mathbb{T}_l$
\STATE Select $\mathcal{N}_{SU}$ reliable unlabeled samples $\boldsymbol{X}_{t,u}^r$ from $\mathbb{T}_u$ using Eq.~27
\STATE Generate pseudo labels $\boldsymbol{Y}_{t,u}^r$ for $\boldsymbol{X}_{t,u}^r$ by $\mathscr{F}^{(r)}(\boldsymbol{\Theta})$
\STATE Update $\mathbb{T}_{l,p}\leftarrow\mathbb{T}_l\cup\{(\boldsymbol{X}_{t,u}^r,\boldsymbol{Y}_{t,u}^r)\}$
\STATE Fine-tune the network $\mathscr{F}^{(r)}(\boldsymbol{\Theta})$ using $\mathbb{T}_{l,p}$
\ENDFOR
\end{algorithmic}
\textbf{End}
\end{algorithm}

\section{Methods}
\subsection{Overall Methods}
Without loss of generality, let us assume a Med-VFM model pre-trained using self-supervised learning on a large number $\mathcal{N}_s$ of unlabeled source-domain images $\boldsymbol{X}_s$ from modality $\mathcal{M}_s$ forming the source domain $\mathbb{S}=\{\boldsymbol{X}_s\}=\{x_1,x_2,...,x_{\mathcal{N}_s}|\mathcal{M}_s\}$. To adapt the Med-VFM on downstream segmentation tasks, the pre-trained Med-VFM is used as an encoder $E$, which is combined with a task-specific decoder $D$ to form a U-shaped segmentation architecture $\mathscr{F}(\boldsymbol{\Theta};[E;D])$. Importantly, the proposed framework is model-agnostic and does not rely on specific architectural assumptions of the underlying Med-VFM. It operates on feature representations extracted from the encoder and can therefore be applied to a wide range of foundation models with different architectures and pre-training strategies. In this work, we demonstrate the approach using a representative Med-VFM for concreteness, but the method is generally applicable across different models and modalities.

The goal is to adapt this segmentation model to a target domain $\mathbb{T}=\{\boldsymbol{X}_t\}=\{x_1,x_2,...,x_{\mathcal{N}_t}|\mathcal{M}_t\}$ consisting of $\mathcal{N}_t$ unlabeled images $\boldsymbol{X}_t$ from modality $\mathcal{M}_t$ without access to the source-domain data. This setting reflects practical deployment scenarios, where only the pre-trained model is available and annotation resources are limited.

To address this challenge, we propose an \textbf{Active Selective Semi-supervised Fine-tuning (ASSFT)} method, which iteratively integrates informed sample selection with selective semi-supervised adaptation. The key idea is to efficiently identify and annotate a small subset of informative samples, while leveraging the remaining unlabeled data through pseudo-labeling to improve segmentation performance under limited annotation budgets.

\textbf{Initialization.} The encoder is initialized from the pre-trained Med-VFM as $E^{(p)}(\boldsymbol{X}_s)$, while the decoder is randomly initialized as $D^{(0)}$. The segmentation model can therefore be represented as $\mathscr{F}(\boldsymbol{\Theta})=\mathscr{F}(\boldsymbol{\Theta};[E^{(p)}(\boldsymbol{X}_s);D^{(0)}])$. Since no labeled data are initially available in the target domain, the dataset is partitioned into an unlabeled set $\mathbb{T}_u=\mathbb{T}$ and an empty labeled set $\mathbb{T}_l=\varnothing$. To bootstrap training, one initial sample $x_1$ is annotated. The annotated pair $(x_1,y_1)$ is added to the labeled target set, and the corresponding image is removed from $\mathbb{T}_u$. The initial model $\mathscr{F}^{(0)}(\boldsymbol{\Theta})$ is then obtained by training on this sample.

\textbf{Active Adaptation Procedure.} As illustrated in Algorithm \ref{alg:algorithm}, ASSFT operates over $\mathcal{R}$ rounds. In each round, a small batch of $\mathcal{N}_B$ informative samples is selected from the unlabeled target set for annotation using the proposed \textbf{Active Test Time Sample Query} strategy. This strategy evaluates candidate samples based on two complementary criteria: \textbf{Diversified Knowledge Divergence (DKD)} and \textbf{Anatomical Segmentation Difficulty (ASD)}. The total number of annotated samples is therefore $\mathcal{N}_{AL}=\mathcal{N}_B\cdot \mathcal{R}$, where $\mathcal{N}_{AL}\ll\mathcal{N}_t$. Starting from round $r=1$, ASSFT alternates between active learning and semi-supervised fine-tuning.
\begin{enumerate}
    \item \textbf{Active Test Time Sample Query.} A batch of $\mathcal{N}_B$ informative samples $\boldsymbol{X}_t^r=\{x_1,...,x_{\mathcal{N}_B}\}$ is selected from the unlabeled set $\mathbb{T}_u$ using the proposed query strategy. These samples are annotated, generating labels $\boldsymbol{Y}_t^r=\{y_1,...,y_{\mathcal{N}_B}\}$. These newly labeled samples are added to the labeled set $\mathbb{T}_l=\mathbb{T}_l\cup\{(\boldsymbol{X}_t^r,\boldsymbol{Y}_t^r)\}$, and removed from the unlabeled set $\mathbb{T}_u=\mathbb{T}_u\backslash\{\boldsymbol{X}_t^r\}$. 
    \item \textbf{Selective Semi-supervised Fine-tuning.} This step consists of three stages:
    \begin{enumerate}[(\roman*)]
    \item Supervised fine-tuning: The segmentation model $\mathscr{F}^{(r)}(\boldsymbol{\Theta})$ is first updated using the labeled target samples $\mathbb{T}_l$ in a fully supervised manner.
    \item Reliable unlabeled sample selection: To further exploit unlabeled data, a subset of $\mathcal{N}_{SU}$ unlabeled samples $\boldsymbol{X}_{t,u}^r$ is selected from $\mathbb{T}_u$ based on predictive confidence and semantic distance. The number of selected samples increases with the iteration number ($\mathcal{N}_{SU}=\mathcal{N}_B\cdot r$), allowing the model to progressively incorporate more unlabeled data as it becomes more reliable.
    \item Pseudo-label-based fine-tuning: Pseudo labels $\boldsymbol{Y}_{t,u}^r$ are then generated by the current model for the selected unlabeled samples, and combined with the labeled set to form an augmented training set $\mathbb{T}_{l,p}=\mathbb{T}_l\cup \{(\boldsymbol{X}_{t,u}^r,\boldsymbol{Y}_{t,u}^r)\}$. The model is then further fine-tuned using this combined dataset.
    \end{enumerate}
\end{enumerate}

This iterative procedure continues until the predefined annotation budget $\mathcal{N}_{AL}$ or the maximum number of rounds $\mathcal{R}$ is reached. Through this process, the segmentation model is progressively refined to obtain the final model $\mathscr{F}(\boldsymbol{\Theta}^*)$, achieving strong segmentation performance on the target domain while requiring only a small number of annotated samples.

\begin{figure*}[!t]
\centering
\includegraphics[width=0.9\textwidth]{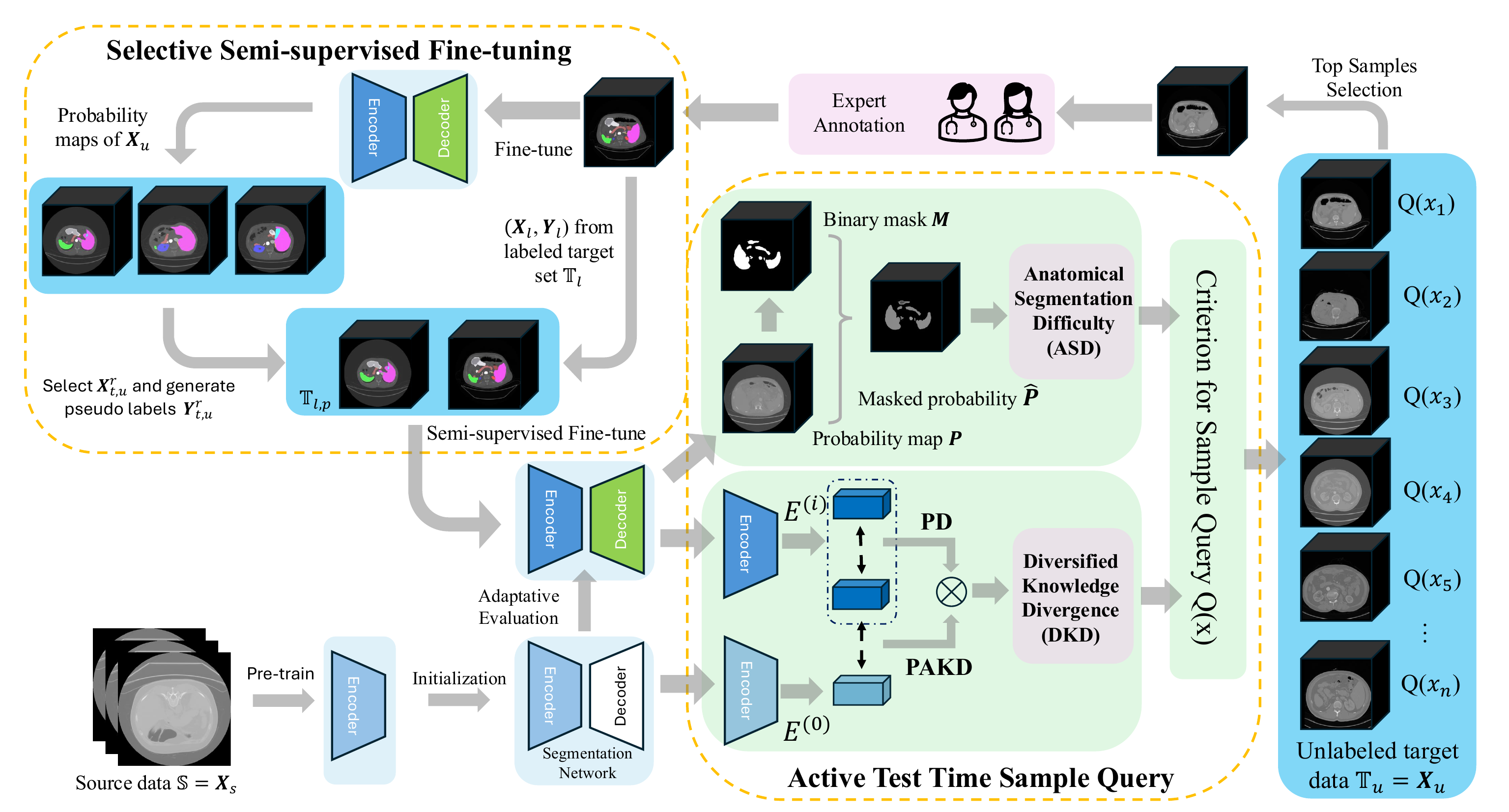}
\caption{\textbf{Active Selective Semi-supervised Fine-tuning} (ASSFT) of medical vision foundation models for volumetric medical image segmentation. The segmentation network was pre-trained on the source data $\mathbb{S}=\{\boldsymbol{X}_s\}$ and adapted to the target domain $\mathbb{T}$ for downstream evaluation. ASSFT employs an \textbf{Active Test Time Sample Query} strategy to evaluate the information level of each target sample. This strategy employs two metrics: \textbf{Diversified Knowledge Divergence} (DKD) and \textbf{Anatomical Segmentation Difficulty} (ASD). The scores of these two metrics are combined to form the criterion for sample query $Q(x)$ of unlabeled target data $\mathbb{T}_u=\{\boldsymbol{X}_u\}$. Top samples $\boldsymbol{X}_l$ with large query scores are selected for annotation $\boldsymbol{Y}_l$ by experts. These queried samples and their annotations are used to fine-tune the network via a \textbf{Selective Semi-supervised Fine-tuning}. After fine-tuned by labeled data $(\boldsymbol{X}_l,\boldsymbol{Y}_l)$, the network makes predictions to generate probability maps for unlabeled target data $\boldsymbol{X}_u$. Unlabeled data $\boldsymbol{X}_{t,u}^r$ are selected and their pseudo labels $\boldsymbol{Y}_{t,u}^r$ are generated. These data $(\boldsymbol{X}_{t,u}^r,\boldsymbol{Y}_{t,u}^r)$ are combined with labeled data $(\boldsymbol{X}_l,\boldsymbol{Y}_l)$ to fine-tune the network.} 
\label{vis1}
\end{figure*}

\subsection{Active Test Time Sample Query}
To guide the selection of informative samples during adaptation, we introduce an Active Test Time Sample Query strategy. The goal of this component is to identify a small subset of target-domain samples whose annotation would most effectively improve the model, particularly in the absence of source-domain data. Because the current setting assumes that only a pre-trained model and unlabeled target data are accessible., sample selection must rely entirely on the model’s current representations and predictions on the target domain. This motivates a test-time querying strategy that evaluates candidate samples based on their potential to refine and adapt the model. To achieve this, we assess sample informativeness from two complementary perspectives: First, we quantify how much a sample deviates from the model’s existing knowledge using Diversified Knowledge Divergence (DKD), which captures both knowledge novelty and diversity in the feature space. Second, we evaluate the difficulty of segmenting anatomically meaningful structures using Anatomical Segmentation Difficulty (ASD), which prioritizes samples that are likely to improve task-specific performance. These two criteria jointly enable the selection of samples that are both informative for updating the model’s representations and relevant for improving segmentation quality. Selected samples are then queried for annotation and incorporated into the subsequent fine-tuning process.

In the following sections, we describe the proposed query metrics in detail. Section 3.2.1 introduces DKD, which focuses on knowledge novelty and diversity, while Section 3.2.2 presents ASD, which captures segmentation difficulty from an anatomical perspective.

\subsubsection{Diversified Knowledge Divergence}
To identify samples that contain informative and diverse target-domain knowledge, we propose a sample-level query metric, termed \textit{Diversified Knowledge Divergence (DKD)}. The goal of DKD is to prioritize samples that both introduce new knowledge relative to the pre-trained model and provide complementary information within the target dataset.  To achieve this, DKD consists of two complementary components: (1) Prior and Adaptive Knowledge Divergence (PAKD), which measures knowledge novelty, and (2) Pair-wise Dissimilarity (PD), which promotes diversity among selected samples.

\textbf{Prior and Adaptive Knowledge Divergence (PAKD).} We first quantify how much a target sample differs from the model’s existing knowledge. Intuitively, if the model’s representation of a sample changes significantly during adaptation, this indicates that the sample contains target-specific information not yet captured by the model.

To measure this, we compare feature embeddings extracted from the initial pre-trained encoder $E^{(0)}$ and the adapted encoder $E^{(i)}$. For an unlabeled target scan $x_u \in \mathbb{R}^{H\times W\times D}$ from $\mathbb{T}_u$, PAKD is defined as the cosine distance ($\textrm{CosDis}$) between these two representations:
\begin{align}
    \nonumber
    \textrm{PAKD}(x_u) &=  \textrm{CosDis}(E^{(0)}(x_u),E^{(i)}(x_u)) \\
    &=1-\frac{E^{(0)}(x_u)\cdot E^{(i)}(x_u)^T}{||E^{(0)}(x_u)||\cdot||E^{(i)}(x_u)||}
\end{align} 
A larger PAKD value indicates a greater discrepancy between the initial and adapted representations, suggesting that the sample contains domain-specific information that has not yet been captured by the model. Thus, selecting samples with high PAKD helps the model better adapt to domain-specific characteristics.

\textbf{Pair-wise Dissimilarity (PD).} While PAKD identifies informative samples, selecting samples solely based on this criterion may lead to redundancy as similar samples may receive similar scores. To mitigate this issue, we introduce a diversity component that encourages selection of samples that are distinct from one another.

All unlabeled samples of the target set $\mathbb{T}_u$ are first ranked in descending order according to their PAKD scores as $x_u^1,x_u^2,\ldots,x_u^{\mathcal{N}_u}$. For a candidate sample $x_u^c$, PD measures its semantic dissimilarity with respect to previously ranked samples using cosine distance between their feature embeddings:
\begin{align}
    \nonumber
    \textrm{PD}(x_u^c)&= \frac{\sum_{k=1}^{c-1} k*\textrm{ConDis}(E^{(i)}(x_u^c),E^{(i)}(x_u^{c-k}))}{\sum_{k=1}^{c-1}k}\\
    &=\frac{\sum_{k=1}^{c-1} k*(1-\frac{E^{(i)}(x_u^c)\cdot E^{(i)}(_u^{c-k})^T}{||E^{(i)}(x_u^c)||\cdot||E^{(i)}(x_u^{c-k})||})}{\sum_{k=1}^{c-1}k}.
\end{align}
The weighting term $k$ emphasizes diversity relative to high-PAKD samples, ensuring that newly selected samples are not only informative but also complementary to already selected ones.

The final DKD score is defined as:
\begin{align}
    \textrm{DKD}(x_u)=\textrm{PAKD}(x_u)\times\textrm{PD}(x_u).
\end{align}
This formulation ensures that selected samples are both: (i) informative, by introducing new knowledge relative to the model, and (ii) diverse, by avoiding redundancy with other selected samples.

\subsubsection{Anatomical Segmentation Difficulty}
In addition to knowledge novelty and diversity, effective sample selection should prioritize samples that are challenging to segment, as these provide valuable supervision for improving task-specific performance. To capture this, we introduce an organ-level query metric, termed \textit{Anatomical Segmentation Difficulty (ASD)}. Unlike conventional uncertainty-based sampling methods that evaluate uncertainty across the entire image, ASD focuses specifically on foreground anatomical regions. This is important in medical imaging, where large background areas can dominate uncertainty estimates and obscure meaningful signals from anatomically relevant structures.

Given an unlabeled sample $x_u \in \mathbb{R}^{H\times W\times D}$ from $\mathbb{T}_u$, the segmentation network $\mathscr{F}(\boldsymbol{\Theta})$ generates $C$ class probability maps:
\begin{align}
    \boldsymbol{P}=\{[\boldsymbol{P}_0,...,...,\boldsymbol{P}_{C-1}]\},
\end{align}
where $\boldsymbol{P}_0$ corresponds to the background class. ASD is computed by measuring the prediction uncertainty within foreground regions, which requires reliable identification of these regions. A key challenge is that, especially in early active learning rounds, the model tends to produce overconfident background predictions, which can suppress true foreground regions. To address this, we introduce a temperature scaling mechanism $\tau(r)$ to reduce the dominance of the background class. The adjusted background probability is defined as:
\begin{align}
    \boldsymbol{P}'_0=\frac{\boldsymbol{P}_0}{\tau(r)},
\end{align}
where the temperature function $\tau(r)$ is dynamically adjusted based on the active learning round $r$:
\begin{align}
\tau(r)=-\frac{3}{2}\frac{\log r}{\log\mathcal{R}}+3.
\end{align}
This schedule gradually decreases the temperature from 3 in the first AL round ($r=1$) to 1.5 in the last round ($r=\mathcal{R}$). 

This dynamic formulation enables the foreground masking process to adapt to the model’s learning stage. In early rounds, when the model is trained with limited supervision, segmentation predictions are often unreliable and exhibit overconfidence in the background class. As a result, voxels belonging to anatomical structures may be incorrectly classified as background, leading to false negatives in the foreground mask and the exclusion of informative regions from ASD computation. To mitigate this issue, a higher temperature value is used to reduce the dominance of the background class, thereby increasing tolerance to potential mis-segmentation and allowing more candidate voxels to be considered as foreground. As the model is progressively refined with additional annotated samples, prediction reliability improves, and the temperature decreases to enable more precise foreground localization. The logarithmic decay schedule ensures a gradual transition between these regimes, reflecting the typically rapid improvement in model performance during early iterations followed by stabilization in later stages.

Using the adjusted background probabilities $\boldsymbol{P}'_0$ and the probabilities of foreground classes $\{\boldsymbol{P}_1,...,\boldsymbol{P}_{C-1}\}$, a binary foreground mask $\boldsymbol{M}\in \boldsymbol{R}^{H\times W\times D}$ is constructed as
\begin{align}
    \boldsymbol{P}_{max}&=\max([\boldsymbol{P}_1,...,\boldsymbol{P}_{C-1}]), \\
    \boldsymbol{M}&=\mathds{1}[\boldsymbol{P}'_0<\boldsymbol{P}_{max}],
\end{align}
where the indicator function selects voxels that are more likely to belong to foreground anatomical structures than to the background.

Then, this mask $\boldsymbol{M}$ is applied to the predicted probability maps to obtain foreground-focused probability maps $\hat{\boldsymbol{P}}\in\mathbb{R}^{C\times H\times W\times D}$: 
\begin{align}
    \hat{\boldsymbol{P}}&=\boldsymbol{P} \odot \boldsymbol{M}.
\end{align}

Lastly, the ASD score for sample $x_u$ is then computed as the entropy of class probabilities within the foreground region:
\begin{align}
    \textrm{ASD}(x_u) =-\sum_{i=0}^{C-1}\hat{\boldsymbol{P}}_i(x_u)\log\hat{\boldsymbol{P}}_i(x_u).
\end{align}

A higher ASD score indicates greater uncertainty in anatomically relevant regions, suggesting that the sample is more challenging and therefore more informative for improving segmentation performance. By focusing on uncertainty within foreground structures, ASD avoids being dominated by background regions and provides a more reliable estimate of task-specific difficulty.

\subsubsection{Unified Sample Query Criterion}
To combine the complementary information captured by DKD and ASD, we define a unified sample selection criterion that integrates both metrics into a single score for ranking candidate samples. Since DKD and ASD are computed on different scales and may exhibit different value distributions, we first normalize each metric using min-max normalization to ensure comparability. For each unlabeled sample $x_u\in\mathbb{T}_u$, the normalized scores are defined as:
\begin{align}
    \tilde{\textrm{DKD}}(x_u)& = \frac{\textrm{DKD}(x_u)-\min(\textrm{DKD}(x_u))}{\max(\textrm{DKD}(x_u))-\min(\textrm{DKD}(x_u))}, \\
    \tilde{\textrm{ASD}}(x_u)& = \frac{\textrm{ASD}(x_u)-\min(\textrm{ASD}(x_u))}{\max(\textrm{ASD}(x_u))-\min(\textrm{ASD}(x_u))}.
\end{align}
While normalization aligns the value ranges, the two metrics may still exhibit skewed or uneven distributions, potentially causing one metric to dominate the combined score. To mitigate this effect, we further apply a quantile transformation, which maps each normalized score to a uniform distribution:
\begin{align}
    \hat{\textrm{DKD}}(x_u)& = \textrm{QuanTrans}(\tilde{\textrm{DKD}}(x_u)), \\
    \hat{\textrm{ASD}}(x_u)& = \textrm{QuanTrans}(\tilde{\textrm{ASD}}(x_u)).
\end{align}
Finally, the unified query criterion $Q$ ($Q\in[0,2]$) is defined as
\begin{align}
    Q(x_u)=\hat{\textrm{DKD}}(x_u)+\hat{\textrm{ASD}}(x_u).
\end{align}

\subsection{Selective Semi-supervised Fine-tuning}
To further improve adaptation to the target domain by enhancing fine-tuning samples while minimizing annotation effort, we propose a Selective Semi-supervised Fine-tuning strategy. The goal of this module is to effectively leverage unlabeled data while controlling the risk of introducing noisy pseudo labels. To this end, we design a three-stage process that first learns from reliable labeled samples, then identifies high-quality unlabeled samples, and finally incorporates them through pseudo-label-based training.

\subsubsection{Stage 1: Supervised Fine-tuning}
This stage establishes a reliable model using the currently available labeled samples. At iteration $r$, the Med-VFM-based segmentation network $\mathscr{F}^{(r)}(\boldsymbol{\Theta})$ is fine-tuned using the labeled target set $\mathbb{T}_l=\{(\boldsymbol{X}_l,\boldsymbol{Y}_l);\mathcal{M}_t\}$. This supervised training allows the networks to progressively capture target-domain characteristics and improves the quality of subsequent predictions on unlabeled samples. 

\subsubsection{Stage 2: Reliable Unlabeled Sample Selection}
This step identifies $\mathcal{N}_{SU}$ unlabeled samples $\boldsymbol{X}_{t,u}^r$ from $\mathbb{T}_u$ for which the model is likely to produce reliable pseudo labels. Selecting such samples is crucial, as inaccurate pseudo labels may degrade model performance. To address this, we evaluate each unlabeled sample using two complementary criteria: (1) predictive confidence and (2) semantic distance to labeled samples.

\paragraph{Predictive Confidence.} Models are more likely to generate accurate segmentation predictions if they demonstrate high predictive confidence, thus producing better pseudo labels. Model confidence is estimated by calculating the difference between the largest and second-largest predicted probabilities in the foreground regions. This margin directly reflects the separation between competing class predictions and provides a stable indicator of prediction reliability.

For an unlabeled sample $x_u\in\mathbb{T}_u$, the largest and second largest predictive probabilities ($\boldsymbol{P}_{max}$ and $\boldsymbol{P}_{max}^{(2)}$) are extracted from predicted probabilities $\boldsymbol{P}_c\in\mathbb{R}^{C\times H\times W\times D}$ as
\begin{align}
    &\boldsymbol{P}_{max}(x_u)=\max\boldsymbol{P}_c(x_u), \\
    &\boldsymbol{P}_{max}^{(2)}(x_u)=\max_{\boldsymbol{P}_c<\boldsymbol{P}_{max}}\boldsymbol{P}_c(x_u).
\end{align}
Since background voxels dominate medical images and usually exhibit high confidence, directly computing confidence over the entire volume may lead to over-estimation. To avoid this issue, we restrict the confidence estimation to predicted foreground regions using a binary mask $\boldsymbol{M}_c\in\mathbb{R}^{H\times W\times D}$, which is designed to separate background regions and foreground structures:
\begin{align}
    \boldsymbol{M}_c(x_u)=\mathds{1}[\textrm{arg}\max_{c\in\{0,1,...,C-1\}}\boldsymbol{P}_c(x_u)\neq0].
\end{align}
The foreground-masked probability maps are then obtained as
\begin{align}
    &\hat{\boldsymbol{P}}_{max}(x_u) =\boldsymbol{P}_{max}(x_u)\odot\boldsymbol{M}_c(x_u), \\
    &\hat{\boldsymbol{P}}_{max}^{(2)}(x_u) =\boldsymbol{P}_{max}^{(2)}(x_u)\odot\boldsymbol{M}_c(x_u).
\end{align}
The overall predictive confidence score for the sample is obtained by aggregating the voxel-wise probability margin over all voxels $v$ in the volume $V$:
\begin{align}
    C(x_u) =\sum_{v\in V} \big(\hat{\boldsymbol{P}}_{max}(x_u)-\hat{\boldsymbol{P}}^{(2)}_{max}(x_u)\big).
\end{align}
To reduce  computational complexity, we pre-select the top $\mathcal{K}$ samples with the highest confidence as candidates samples $\boldsymbol{X}_k$. The candidate size $\mathcal{K}$ is dynamically adjusted based on the model's average confidence across the unlabeled set $\mathbb{T}_u$:
\begin{align}
    \bar{C}=\frac{\sum_{x_u\in\boldsymbol{X}_u}C(x_u)}{U},
\end{align}
where $U$ is the number of samples from the the unlabeled target set $\mathbb{T}_u$. Thus, the number of candidates $\mathcal{K}$ is determined by
\begin{align}
    \mathcal{K}=\frac{\mathcal{N}_{SU}}{\bar{C}/\tau_c}.
\end{align}
where $\tau_c$ is a temperature parameter used to moderate the confidence scale.

This adaptive strategy reflects the model’s learning status: when the model exhibits high average confidence, its predictions are more reliable, and a smaller set of high-confidence samples is sufficient for candidate selection. In contrast, when confidence is low, pseudo-label reliability is reduced, and a larger candidate pool is retained to avoid overlooking informative samples. Consequently, the candidate size is set inversely proportional to the average predictive confidence, enabling an effective balance between computational efficiency and robust sample selection.

\paragraph{Semantic Distance.} To further ensure the reliability of pseudo labels, we evaluate the semantic similarity between candidate samples and labeled samples in the feature space. The intuition is that if an unlabeled sample is similar to labeled ones, the model is more likely to generate accurate predictions. Feature embeddings of labeled samples $x_l\in\mathbb{T}_l$ are first extracted by the encoder $E^{(r)}$ as anchors $\boldsymbol{A}^{(r)}$ for candidate samples:
\begin{align}
    \boldsymbol{A}^{(r)}(x_l)=E^{(r)}(x_l).
\end{align}
Similarly, feature embeddings for candidate samples $\boldsymbol{X}_k=\{x_k|k\in\{1,...,{\mathcal{K}}\}\}$ are obtained as 
\begin{align}
    \boldsymbol{F}^{(r)}(x_k)=E^{(r)}(x_k).
\end{align}
The semantic distance $D$ of each candidate sample $x_k$ is defined as the minimum cosine distance to all anchors $\boldsymbol{A}^{(r)}$:
\begin{align}
    D(x_k) =  \min_{a\in\boldsymbol{A}^{(r)}}\big[\textrm{CosDis}\big(\boldsymbol{A}^{(r)},\boldsymbol{F}^{(r)}(x_k)\big)\big].
\end{align} 
Finally, the reliability score $R$ of a candidate sample $x_k$ is computed by combining predictive confidence and semantic similarity as
\begin{align}
    R(x_k)=C(x_k)\times(1-D(x_k)).
\end{align}

Unlabeled samples with the highest reliability scores are selected for pseudo labeling. To maintain balance between supervised and pseudo-supervised training, the number of selected unlabeled samples is set equal to the number of labeled samples in the current iteration, i.e., $\mathcal{N}_{SU}=\mathcal{N}_B\cdot r$.

\subsubsection{Stage 3: Pseudo-label-based Fine-tuning}
In the final stage, pseudo labels $\boldsymbol{Y}_{t,u}$ are generated for the selected unlabeled samples $\boldsymbol{X}_{t,u}$ using the current segmentation model. These pseudo-labeled samples $(\boldsymbol{X}_{t,u},\boldsymbol{Y}_{t,u})$ are then combined with the labeled set $\mathbb{T}_l$ to form an augmented training set $\mathbb{T}_{l,p}=\mathbb{T}_l\cup\{(\boldsymbol{X}_{t,u}^r, \boldsymbol{Y}_{t,u}^r); \mathcal{M}_t\}$. The segmentation network is subsequently fine-tuned using this combined dataset in a semi-supervised manner, allowing it to exploit additional unlabeled data while maintaining robustness to potential noise in pseudo labels.

\subsection{Exclusion of Redundant Samples from Querying}
During Selective Semi-supervised Fine-tuning, $\mathcal{N}_{SU}$ unlabeled samples $\boldsymbol{X}_{t,u}$ are selected for pseudo labeling based on small semantic distance to labeled samples and high predictive confidence. These characteristics indicate that their semantic patterns have already been well captured by the current model. Consequently, these samples would provide limited additional information if selected for manual annotation, as their inclusion would largely reinforce already learned representations rather than introduce new knowledge.

To avoid redundant annotation and improve the efficiency of active learning, these pseudo-labeled samples are excluded from the candidate pool for active sample querying in the subsequent AL round. This ensures that annotation efforts are focused on samples that are both informative and complementary to the existing labeled set.

\subsection{The Architecture of Med-VFM-based Segmentation Network}
The segmentation network is implemented as an U-shape encoder-decoder architecture where the encoder is initialized from CT-FM \citep{pai2025vision}, and the decoder is task specific. The network consists of five resolution levels, including a bottleneck layer, with feature channel sizes of 32, 64, 256, 512, and 1024, respectively. At each level of both the encoder and decoder, feature extraction is performed using residual segmentation blocks. The numbers of residual blocks across the five levels are 1, 2, 2, 4, and 4, respectively. Each residual block contains two consecutive convolutional layers, each followed by batch normalization and a ReLU activation function.

In the encoder, spatial resolution is progressively reduced using $3\times3\times3$ convolutional layers with a stride of 2. In the decoder, feature maps are upsampled using $3\times3\times3$ transposed convolutional layers. Skip connections are applied between corresponding encoder and decoder levels, where encoder features are fused with the upsampled decoder features via channel-wise addition.

At the input stage, a $3\times3\times3$ convolutional layer maps the input channels to 32 feature channels. At the output stage, a $1\times1\times1$ convolutional layer produces the final segmentation prediction. The encoder is initialized with weights pre-trained on large-scale CT datasets using self-supervised learning as described in CT-FM \citep{pai2025vision}, while the decoder is randomly initialized using Kaiming initialization \citep{he2015delving}.

\section{Experiments}
\subsection{Datasets}
To evaluate the effectiveness of our methods, we adapted the Med-VFM-based segmentation network to five abdominal multi-organ segmentation datasets with large differences in the number of subjects, imaging modalities, and the number of organs to be segmented.

\textbf{AMOS2022-CT}. The AMOS2022-CT dataset was released as part of the MICCAI 2022 AMOS abdominal multi-organ segmentation challenge \citep{ji2022amos}. It consists of 300 abdominal CT volume scans with voxel-level annotations for 15 organs, including the spleen, right kidney, left kidney, gallbladder, esophagus, liver, stomach, aorta, inferior vena cava, pancreas, right adrenal gland, left adrenal gland, duodenum, bladder, and prostate/uterus. Each CT volume consists of $67 \sim 369$ slices of $512 \times 512$ pixels with a slice spacing of $1.25 \sim 5.00$ mm.

\textbf{FLARE 2021}. This dataset was released as part of The Fast and Low GPU memory Abdominal oRgan sEgmentation (FLARE) challenge \citep{ma2022fast}. It consists of 361 CT images with voxel-wise annotations for four abdominal organs, including the liver, the kidneys, the spleen, and the pancreas. It exhibits substantial diversity across various centers, vendors, phases, and diseases. 

\textbf{Abdomen Atlas}. This dataset includes 9,262 3D CT volumes with annotations for 25 anatomical structures, including 16 abdominal organs, two thoracic organs, five vascular structures, and two skeletal structures \citep{li2025well}. The data were collected from 112 hospitals in 19 countries, resulting in a highly heterogeneous, multi-center dataset.

\textbf{AMOS2022-MRI}. This dataset was released as part of the MICCAI 2022 AMOS abdominal multi-organ segmentation challenge \citep{ji2022amos}. It consists of 60 abdominal MR volume scans with voxel-level annotations for 13 organ categories, including the spleen, right kidney, left kidney, gallbladder, esophagus, liver, stomach, aorta, inferior vena cava, pancreas, right adrenal gland, left adrenal gland, and duodenum. Since only two cases have annotations of bladder and prostate/uterus, these two labels were excluded from this dataset to avoid extreme class imbalance.

\textbf{Abdominal MRI}. This dataset was derived from the external evaluation of MRAnnotator \citep{zhou2025mrannotator}. We selected 30 abdominal MR images from external evaluation with voxel-wise annotations for eight anatomical structures, including the spleen, right kidney, left kidney, gallbladder, liver, aorta, inferior vena cava, and pancreas.

Scans from all datasets were preprocessed using a standard pipeline. Specifically, the image intensities were clipped at the 5th and 95th percentiles, followed by z-score normalization applied to each volume. Subsequently, scans were cropped to sub-volumes of fixed size. The patch size was $96\times160\times160$ in the AMOS2022-CT dataset, $96\times192\times160$ in the FLARE 2021 dataset, $96\times128\times160$ in the Abdomen Atlas dataset, $48\times160\times224$ in the AMOS2022-MRI dataset, and $64\times192\times192$ in the Abdominal MRI dataset.

\subsection{Implementation Details}
The experiments were implemented using PyTorch\footnote{http://pytorch.org/}. A combination of dice loss $\mathcal{L}_{Dice}$ and cross-entropy loss $\mathcal{L}_{CE}$ was used as the loss function. The loss function $\mathcal{L}$, defined by the prediction $\hat{y}$ and the ground truth $y$, is given by
\begin{align}
    \nonumber
    \mathcal{L}=\mathcal{L}_{Dice}(\hat{y},y)+\mathcal{L}_{CE}(\hat{y},y).
\end{align}

Models were trained with a batch size of 2 on an NVIDIA Tesla A100 PCI-E Passive Single GPU with 40GB of GDDR5 memory. The AdamW optimizer was used. The network was trained for 100 epochs in one-shot initialization. Subsequently, the network was fined-tuned for 800 epochs per round in the Abdomen Atlas domain, and 400 epochs per round in the remaining four domains. The learning rate was initially set to $0.0001$, and it decayed with a polynomial learning rate scheduler. During training, data augmentation techniques was applied to improve model robustness. Specifically, patches were rotated between $[-30, 30]$ along each axis with a probability of $0.2$ and then scaled between $(0.7,1.4)$ with a probability of $0.2$. Subsequently, all patches were mirrored along all axes with a probability of $0.5$. Zero-centered additive Gaussian noise, with variance drawn from the distribution $U(0, 0.1)$, and brightness adjustments were added to each voxel sample, each with a probability of $0.15$. Segmentation performance was evaluated using the dice coefficient score (Dice).

\subsection{Active Learning Protocols}
We implemented different active learning protocols for different datasets to account for differences in segmentation complexity and imaging modalities. For the AMOS2022-CT, FLARE 2021, and AMOS2022-MRI datasets, we implemented an active domain adaptation strategy, in which samples were iteratively queried to fine-tune the segmentation network on the target domain. Specifically, the AL procedure in AMOS2022-CT and FLARE 2021 was conducted for five rounds ($\mathcal{R}=5$), with a query budget of $5\%$ per round ($\mathcal{N}_B=5\%$). For AMOS2022-MRI, we used six rounds with the same query budget ($\mathcal{N}_B=5\%$) per round, reflecting the greater domain shift between pretraining and target data due to differing imaging modalities. Additionally, we reported results for two reference settings: a lower bound ($0\%$), corresponding to source-only training without target-domain fine-tuning, and an upper bound ($100\%$), corresponding to fully supervised training using all target-domain samples.

For the Abdomen Atlas and Abdominal MRI datasets, we employed an active few-shot learning strategy, querying a small number of samples for fine-tuning. In Abdomen Atlas, we queried $5\%$ ($\mathcal{N}_B=5\%$) and $10\%$ ($\mathcal{N}_B=10\%$) samples, respectively. In the Abdominal MRI dataset, we queried $2$ samples ($\mathcal{N}_B=2$) per round for three rounds, corresponding active 3-shot learning, active 5-shot learning and active 7-shot learning, respectively. 

\begin{table*}[!t]
\centering
\caption{Performance comparison between our ASSFT, DKD+ASD, and other SOTA methods when adapting Med-VFMs to the AMOS2022-CT domain. The performance was evaluated using Dice scores, and the results were reported as Mean$\pm$SD. \textbf{Bold} and \underline{underline} represent the best and the second best results. Our ASSFT method achieved superior performance than other AL and ADA methods (Lower bound $0\%$: source-training without target fine-tuning; Upper bound $100\%$: fully supervised training by all target samples; $^*$: $p<0.01$ with the Mann-Whitney U test between ASSFT and SOTA methods).}
\label{tab1}
\resizebox{\textwidth}{!}{
\begin{tabular}{c|c|c|c|c|c|c|c}
\toprule
\multirow{2}{*}{Methods} & Lower Bound & \multicolumn{5}{c|}{Query Budgets (Number of Iterations)} & Upper Bound \\
\cline{3-7}
 &  $0\%$ & $5\%$ (r=1) & $10\%$ (r=2) & $15\%$ (r=3) & $20\%$ (r=4) & $25\%$ (r=5) & $100\%$ \\
\midrule
RAND     & 0.27$\pm$0.10   & 73.31$\pm$15.50 & 77.57$\pm$14.65 & 79.30$\pm$14.40 & 81.63$\pm$12.87 
         & 82.30$\pm$12.50 & 93.04$\pm$1.23 \\
ENPY     & 0.27$\pm$0.10   & 74.24$\pm$14.53 & 78.17$\pm$13.73 & 80.78$\pm$13.43 & 82.59$\pm$13.63 
         & 83.31$\pm$12.05 & 93.04$\pm$1.23 \\
LCON     & 0.27$\pm$0.10   & 73.39$\pm$14.75 & 77.62$\pm$14.00 & 80.11$\pm$13.74 & 82.14$\pm$12.13 
         & 82.34$\pm$12.36 & 93.04$\pm$1.23 \\
MMAR     & 0.27$\pm$0.10   & 73.92$\pm$14.99 & 77.63$\pm$13.89 & 80.44$\pm$13.05 & 82.23$\pm$13.00 
         & 82.82$\pm$12.00 & 93.04$\pm$1.23 \\
Core-set & 0.27$\pm$0.10   & 74.42$\pm$14.59 & 78.55$\pm$13.16 & 81.11$\pm$12.45 & 82.73$\pm$12.15 
         & 84.04$\pm$11.54 & 93.04$\pm$1.23 \\
BADGE    & 0.27$\pm$0.10   & 75.78$\pm$14.28 & 79.67$\pm$12.94 & 82.09$\pm$12.50 & 83.56$\pm$11.95 
         & 84.69$\pm$10.17 & 93.04$\pm$1.23 \\
SANN     & 0.27$\pm$0.10   & 75.48$\pm$14.17 & 78.99$\pm$12.43 & 81.88$\pm$11.17 & 83.11$\pm$10.89 
         & 84.45$\pm$10.43 & 93.04$\pm$1.23 \\
UGTST    & 0.27$\pm$0.10   & 75.80$\pm$14.06 & 80.59$\pm$12.92 & 82.22$\pm$11.72 & 83.93$\pm$10.79 
         & 85.27$\pm$10.72 & 93.04$\pm$1.23 \\
CUP      & 0.27$\pm$0.10   & 73.92$\pm$14.85 & 77.88$\pm$13.49 & 80.69$\pm$13.05 & 82.25$\pm$12.59 
         & 83.10$\pm$11.70 & 93.04$\pm$1.23 \\
\midrule
DKD+ASD     & 0.27$\pm$0.10 & \underline{78.87}$\pm$13.92 & \underline{82.61}$\pm$12.51 & \underline{84.03}$\pm$10.48 & \underline{85.54}$\pm$9.74 
            & \underline{86.78}$\pm$9.57 & 93.04$\pm$1.23 \\
ASSFT       & 0.27$\pm$0.10 & \textbf{80.51}$^*\pm$12.53 & \textbf{84.42}$^*\pm$11.65 & \textbf{85.52}$^*\pm$10.06 & \textbf{86.63}$^*\pm$9.28 
            & \textbf{87.68}$^*\pm$9.08 & 93.04$\pm$1.23 \\
\bottomrule
\end{tabular}}
\end{table*}

\begin{figure*}[!t]
\centering
\includegraphics[width=0.9\textwidth]{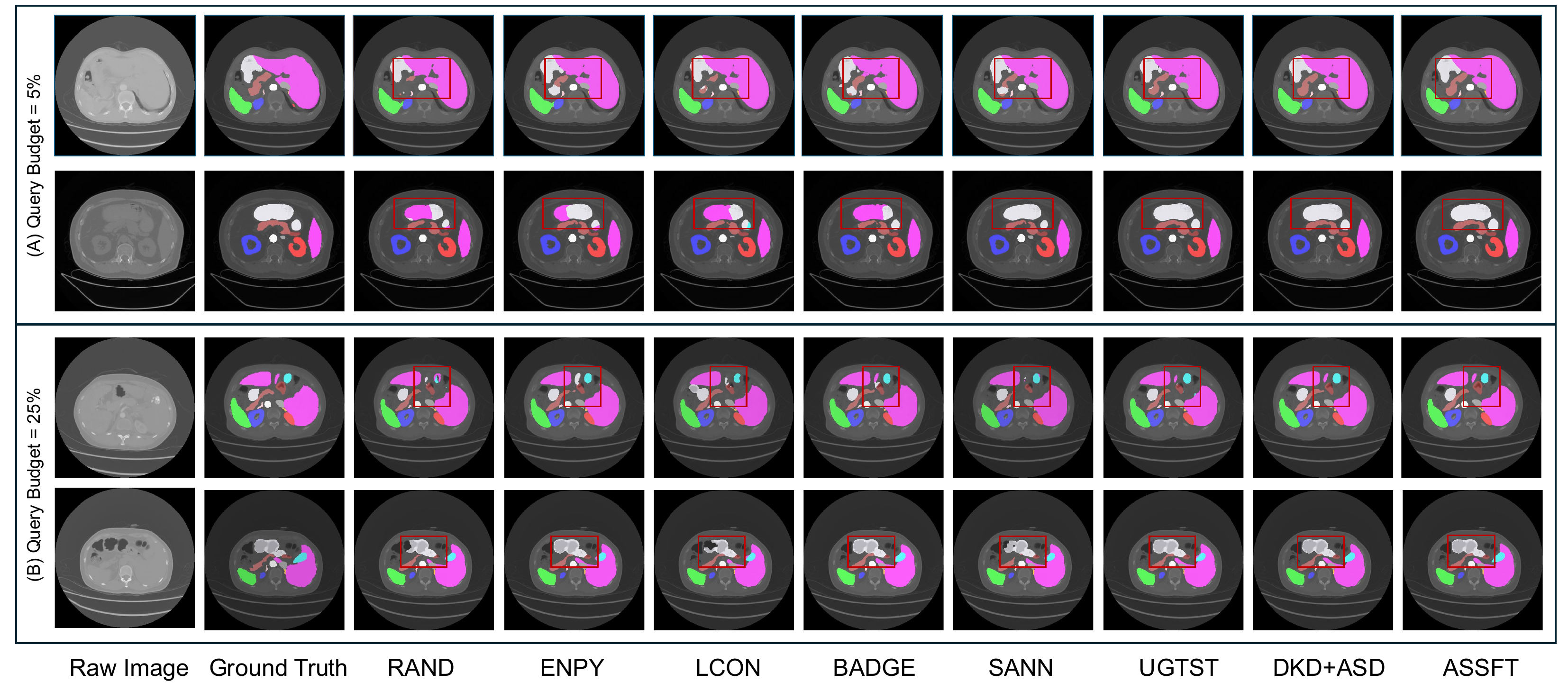}
\caption{Qualitative comparison among results of the medical vision foundation models fine-tuned by (A) $5\%$ and (B) $25\%$ samples from the AMOS2022-CT domain queried by our methods and other SOTA methods. Red boxes mark the regions where our methods exhibit better segmentation results than SOTA methods.} 
\label{vis2}
\end{figure*}

\begin{table*}[!t]
\centering
\caption{Performance comparison between our ASSFT, DKD+ASD,  and other SOTA methods when adapting Med-VFMs to the AMOS2022-MRI domain. The performance was evaluated using Dice scores, and the results were reported as Mean$\pm$SD. \textbf{Bold} and \underline{underline} represent the best and the second best results. Our ASSFT method achieved superior performance than other AL and ADA methods (Lower bound $0\%$: source-training without target fine-tuning; Upper bound $100\%$: fully supervised training by all target samples; $^*$: $p<0.01$ with the Mann-Whitney U test between ASSFT and SOTA methods).}
\label{tab2}
\resizebox{\textwidth}{!}{
\begin{tabular}{c|c|c|c|c|c|c|c|c}
\toprule
\multirow{2}{*}{Methods} & Lower Bound & \multicolumn{6}{c|}{Query Budgets (Number of Iterations)} & Upper Bound \\
\cline{3-8}
 &  $0\%$ & $5\%$ (r=1) & $10\%$ (r=2) & $15\%$ (r=3) & $20\%$ (r=4) & $25\%$ (r=5) & $30\%$ (r=6) & $100\%$ \\
\midrule
RAND     & 0.46$\pm$0.13   & 35.84$\pm$27.14 & 48.06$\pm$25.24 & 55.28$\pm$21.02 
         & 69.15$\pm$14.46 & 75.56$\pm$12.70 & 77.98$\pm$11.88 & 91.32$\pm$3.74 \\
ENPY     & 0.46$\pm$0.13   & 40.43$\pm$23.29 & 58.74$\pm$19.88 & 71.16$\pm$14.35 
         & 76.65$\pm$14.07 & 78.09$\pm$11.85 & 80.94$\pm$9.70  & 91.32$\pm$3.74 \\
LCON     & 0.46$\pm$0.13   & 38.05$\pm$25.62 & 52.12$\pm$24.57 & 62.35$\pm$18.62 
         & 73.78$\pm$13.79 & 76.12$\pm$12.44 & 79.29$\pm$11.72 & 91.32$\pm$3.74 \\
MMAR     & 0.46$\pm$0.13   & 38.95$\pm$25.68 & 55.37$\pm$23.76 & 63.92$\pm$18.67 
         & 75.79$\pm$13.93 & 77.20$\pm$12.11 & 80.61$\pm$10.94 & 91.32$\pm$3.74 \\
Core-set & 0.46$\pm$0.13   & 41.88$\pm$21.56 & 63.04$\pm$22.01 & 72.59$\pm$15.91 
         & 77.24$\pm$11.02 & 78.89$\pm$10.03 & 82.48$\pm$9.18  & 91.32$\pm$3.74 \\
BADGE    & 0.46$\pm$0.13   & 44.00$\pm$21.99 & 65.49$\pm$19.78 & 75.65$\pm$13.13 
         & 79.30$\pm$10.98 & 81.31$\pm$10.89 & 82.80$\pm$8.42  & 91.32$\pm$3.74 \\
SANN     & 0.46$\pm$0.13   & 42.03$\pm$24.73 & 63.13$\pm$22.63 & 74.34$\pm$13.42 
         & 78.68$\pm$10.70 & 79.19$\pm$10.74 & 82.79$\pm$9.54  & 91.32$\pm$3.74 \\
UGTST    & 0.46$\pm$0.13   & 45.29$\pm$21.30 & 66.14$\pm$19.20 & 76.45$\pm$12.32 
         & 80.86$\pm$10.52 & 81.86$\pm$10.49 & 83.53$\pm$8.38  & 91.32$\pm$3.74 \\
CUP      & 0.46$\pm$0.13   & 41.49$\pm$24.83 & 59.57$\pm$20.87 & 72.17$\pm$14.89 
         & 77.14$\pm$12.85 & 78.30$\pm$11.17 & 81.62$\pm$9.12  & 91.32$\pm$3.74 \\
\midrule
DKD+ASD  & 0.46$\pm$0.13   & \underline{49.31}$\pm$20.59 & \underline{69.66}$\pm$18.31 & \underline{79.52}$\pm$11.08 
         & \underline{83.37}$\pm$10.17 & \underline{84.25}$\pm$9.91  & \underline{85.18}$\pm$8.28  & 91.32$\pm$3.74 \\
ASSFT    & 0.46$\pm$0.13   & \textbf{52.06}$^*\pm$18.12 & \textbf{72.06}$^*\pm$17.70 & \textbf{81.59}$^*\pm$10.39 
         & \textbf{85.74}$^*\pm$9.98  & \textbf{86.13}$^*\pm$9.47  & \textbf{86.90}$^*\pm$7.67  & 91.32$\pm$3.74 \\
\bottomrule
\end{tabular}}
\end{table*}

\begin{figure*}[!t]
\centering
\includegraphics[width=0.9\textwidth]{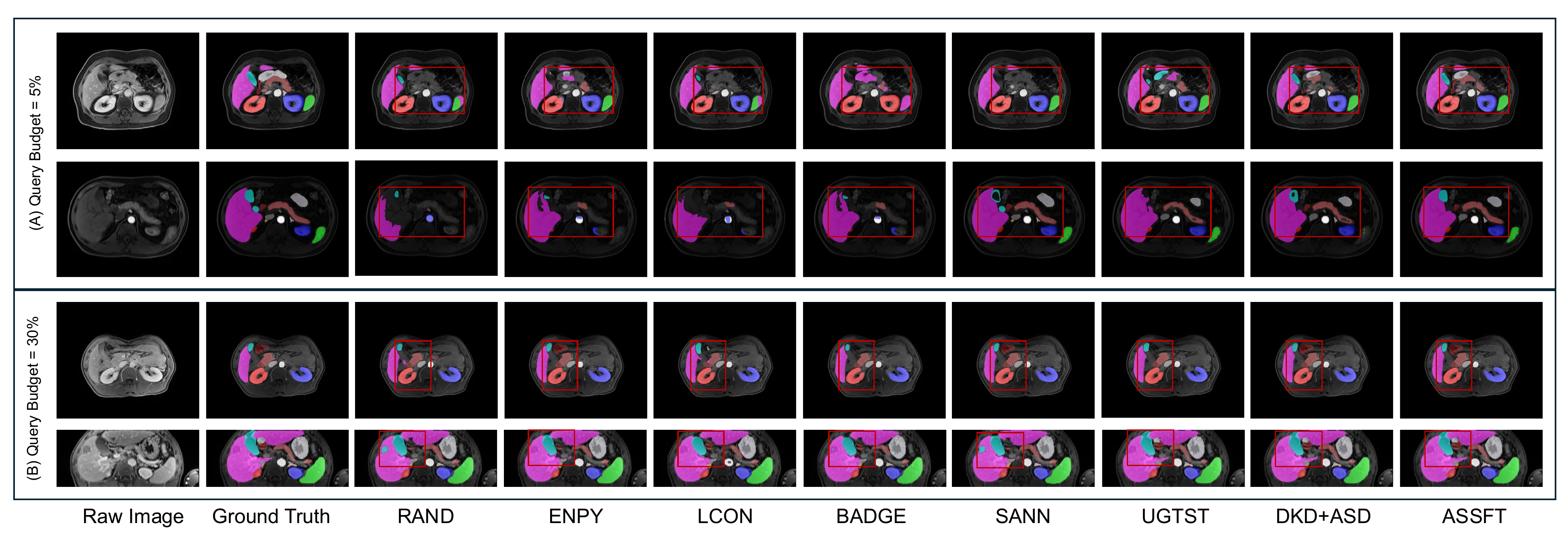}
\caption{Qualitative comparison among results of the medical vision foundation models fine-tuned by (A) $5\%$ and (B) $30\%$ samples from the AMOS2022-MRI domain queried by our methods and other SOTA methods. Red boxes mark the regions where our methods exhibit better segmentation results than SOTA methods.} 
\label{vis3}
\end{figure*}

\begin{table*}
\centering
\caption{Performance comparison between our ASSFT, DKD+ASD, and other SOTA methods when adapting Med-VFMs to the FLARE 2021 domain. The performance was evaluated using Dice scores, and the results were reported as Mean$\pm$SD. \textbf{Bold} and \underline{underline} represent the best and the second best results. Our ASSFT method achieved superior performance than other AL and ADA methods (Lower bound $0\%$: source-training without target fine-tuning; Upper bound $100\%$: fully supervised training by all target samples; $^*$: $p<0.01$ with the Mann-Whitney U test between ASSFT and SOTA methods).}
\label{tab3}
\resizebox{\textwidth}{!}{
\begin{tabular}{c|c|c|c|c|c|c|c}
\toprule
\multirow{2}{*}{Methods} & Lower Bound & \multicolumn{5}{c|}{Query Budgets (Number of Iterations)} & Upper Bound \\
\cline{3-7}
 &  $0\%$ & $5\%$ (r=1) & $10\%$ (r=2) & $15\%$ (r=3) & $20\%$ (r=4) & $25\%$ (r=5) & $100\%$ \\
\midrule
RAND     & 5.86$\pm$0.02  & 86.70$\pm$9.93 & 89.74$\pm$6.23 & 90.57$\pm$5.52 & 91.56$\pm$5.19 & 92.05$\pm$4.18 
         & 96.85$\pm$1.06 \\
ENPY     & 5.86$\pm$0.02  & 88.71$\pm$7.92 & 91.54$\pm$4.65 & 92.07$\pm$4.05 & 92.26$\pm$3.91 & 93.35$\pm$3.25
         & 96.85$\pm$1.06 \\
LCON     & 5.86$\pm$0.02  & 89.70$\pm$6.49 & 92.05$\pm$5.11 & 92.91$\pm$3.35 & 93.74$\pm$2.96 & 93.83$\pm$2.62 
         & 96.85$\pm$1.06 \\
MMAR     & 5.86$\pm$0.02  & 89.68$\pm$5.39 & 92.08$\pm$4.21 & 92.71$\pm$3.18 & 93.50$\pm$2.66 & 93.84$\pm$2.58 
         & 96.85$\pm$1.06 \\
Core-set & 5.86$\pm$0.02  & 89.16$\pm$6.46 & 91.69$\pm$4.23 & 92.61$\pm$3.75 & 92.85$\pm$3.05 & 93.36$\pm$3.19 
         & 96.85$\pm$1.06 \\
BADGE    & 5.86$\pm$0.02  & 89.51$\pm$5.69 & 91.77$\pm$4.13 & 92.65$\pm$3.85 & 92.99$\pm$3.53 & 93.80$\pm$2.76
         & 96.85$\pm$1.06 \\
SANN     & 5.86$\pm$0.02  & 89.60$\pm$5.44 & 92.30$\pm$3.80 & 93.43$\pm$3.15 & 93.75$\pm$2.92 & 93.98$\pm$2.91
         & 96.85$\pm$1.06 \\
UGTST    & 5.86$\pm$0.02  & 89.56$\pm$5.88 & 92.15$\pm$3.87 & 93.26$\pm$3.17 & 93.60$\pm$2.67 & 93.70$\pm$3.05 
         & 96.85$\pm$1.06 \\
CUP      & 5.86$\pm$0.02  & 88.29$\pm$8.87 & 90.47$\pm$5.39 & 91.59$\pm$4.38 & 91.90$\pm$4.20 & 92.31$\pm$4.12 
         & 96.85$\pm$1.06 \\
\midrule
DKD+ASD     & 5.86$\pm$0.02 & \underline{91.65}$\pm$5.09 & \underline{93.30}$\pm$3.68 & \underline{93.94}$\pm$3.08 & \underline{94.46}$\pm$2.78 & \underline{94.60}$\pm$2.53 
            & 96.85$\pm$1.06 \\
ASSFT       & 5.86$\pm$0.02 & \textbf{92.52}$^*\pm$4.31 & \textbf{94.08}$^*\pm$3.27 & \textbf{94.33}$^*\pm$3.04 & \textbf{94.54}$^*\pm$2.64 & \textbf{94.87}$^*\pm$2.47 
            & 96.85$\pm$1.06 \\
\bottomrule
\end{tabular}}
\end{table*}

\subsection{Comparison with State-of-the-art Methods}
To comprehensively evaluate the effectiveness of the proposed method, we implemented two variants. First, the full \textbf{ASSFT} framework was applied to adapt the segmentation network following Algorithm~\ref{alg:algorithm}, incorporating both the Active Test-Time Sample Query strategy and the Selective Semi-supervised Fine-tuning strategy. Second, we implemented a simplified variant, denoted as \textbf{DKD+ASD}, which utilizes the proposed Active Test-Time Sample Query strategy (based on both DKD and ASD) to select samples but performs fine-tuning using only fully supervised learning without the Selective Semi-supervised Fine-tuning component.

To provide a rigorous and comprehensive evaluation, we compared our approach against both established baseline methods and recent approaches specifically designed for medical image segmentation. We first included several commonly used AL or ADA strategies. These are standard and widely adopted baselines in the literature, which enable consistent benchmarking and facilitate comparison with prior work. These include: (1) \textbf{Random Selection (RAND)}, which randomly selects samples for annotation;  (2) \textbf{Entropy (ENPY)}: an uncertainty-based method that queries samples with the highest predictive entropy \citep{wang2014new}; (3) \textbf{Least Confidence (LCON)}, which selects samples with the lowest prediction confidence \citep{li2006confidence}; (4) \textbf{Minimal Margin (MMAR)}, which selects samples with the smallest margin between the top two predicted class probabilities \citep{wang2014new}; (5) \textbf{Core-set}, which is a diversity-based method to select representative samples by solving a set-cover problem \citep{sener2018active}; (6) \textbf{Batch Active learning by Diverse Gradient Embeddings (BADGE)}: which is a hybrid uncertainty- and diversity-based strategy to select samples from diverse batches constructed by calculating gradients and running K-Means++ \citep{Ash2020Deep}. Additionally, we compared our methods with AL methods specifically designed for medical image segmentation,. In particular, we compared with \textbf{Suggestive Annotations (SANN)}, a hybrid uncertainty- and diversity-based method that selects samples with high predictive uncertainty and low semantic similarity \citep{yang2017suggestive}. Importantly, we further compared our approach with recent state-of-the-art (SOTA) active source-free domain adaptation methods proposed for medical image segmentation, which represent the most directly relevant class of approaches for this task. These include: (1) \textbf{Uncertainty-guided Tiered Self-training (UGTST)}, which selects samples with large aleatoric uncertainty and low neighboring diversity in MR prostate segmentation \citep{luo2024uncertainty}; and (2) \textbf{Cascade Uncertainty Predominance (CUP)}, which queries image patches with low prediction uncertainty and large foreground regions for retinal vessel segmentation \citep{wang2024advancing}.

\subsubsection{Results on AMOS2022-CT}
On the AMOS2022-CT dataset, both ASSFT and the proposed query strategy (DKD+ASD) consistently achieved the best and second-best Dice scores across all query budgets (Table~\ref{tab1}). With $5\%$–$25\%$ of samples queried, ASSFT achieved 5–8 Dice points higher performance than random selection. With only $5\%$ queried samples, ASSFT achieved a Dice score of $80.51$, outperforming the strong baseline UGTST and SANN by approximately 4.7 and 5.0 points, respectively. Similar performance gaps were maintained as the query budget increased. DKD+ASD achieved 3–4 Dice points higher scores than UGTST and SANN when $5\%$ and $10\%$ of samples were queried, and maintained 1.5–3 points improvements when $25\%$ samples were queried. These results indicated that the proposed query strategy effectively selected informative samples and that the ASSFT fine-tuning framework further improved adaptation performance. Qualitative comparison provided more evidence to demonstrate the superior performance of ASSFT compared to other approaches (Figure~\ref{vis2}). Segmentation models adapted by $5\%$ ASSFT-queried samples correctly segmented target anatomical structures, especially organs with large varying shapes and sizes, such as pancreas and stomach, while models adapted by other approaches mis-segmented these organs or mis-assigned neighboring segmentation labels to them.

\subsubsection{Results on AMOS2022-MRI}
On the AMOS2022-MRI dataset, ASSFT achieved the best performance across all query budgets compared to SOTA methods (Table \ref{tab2}). With $5\%$ queried samples, ASSFT outperformed random selection and UGTST by approximately 16 and 6.8 Dice points, respectively. With $30\%$ queried samples, ASSFT achieved 9 Dice points higher than random querying. DKD+ASD consistently ranked second among all query strategies, demonstrating the effectiveness of combining the DKD and ASD metrics for informative sample selection. When querying $5\%$ and $30\%$ of samples for supervised fine-tuning, DKD+ASD improved Dice scores by more than 13 and 7 points compared with RAND, respectively. Moreover, DKD+ASD achieved over 5 Dice points improvement compared with BADGE and about 4 points improvement over UGTST when only $5\%$ samples were queried. Qualitative comparison provided more evidence to demonstrate the superior performance of ASSFT on cross-modality adaptation compared to other approaches (Figure~\ref{vis3}). CT-trained segmentation models adapted by $5\%$ ASSFT-queried MR samples correctly segmented target anatomical structures, while other approaches were unable to adapt models to recognize target anatomical structures, thus mis-segmenting them. Additionally, selecting $30\%$ samples by ASSFT adapted models to correctly segment target anatomical structures, but models adapted by other approaches under-segmented those organs.

\subsubsection{Results on FLARE2021}
On the FLARE2021 dataset, ASSFT achieved superior performance compared with other AL and ADA methods (Table~\ref{tab3}). With $5\%$ queried samples, ASSFT achieved $92.52$ Dice points, outperforming SANN and UGTST by about 3 points. With $5\%$–$25\%$ queried samples, ASSFT consistently outperformed RAND, LCON, and MMAR by approximately 1–6 Dice points. DKD+ASD also outperformed all competing AL and ADA methods across different query budgets. Fine-tuning with $5\%$–$25\%$ samples queried by DKD+ASD yielded 1–2 Dice points higher scores than SANN and UGTST. The consistent improvement across query iterations further demonstrates the robustness of the proposed query strategy and ASSFT framework. Qualitative results demonstrated that selecting $5\%$ samples by ASSFT adapted models to generate precise segmentation masks for anatomical structures with varying shapes and sizes (e.g., pancreas), while other methods did not adapt models to achieve accurate segmentation in this dataset (Figure~\ref{vis4}).

\subsubsection{Results on Abdomen Atlas}
On the Abdomen Atlas dataset, ASSFT achieved the highest Dice scores among all AL and ADA methods (Table~\ref{tab4}). With $5\%$ and $10\%$ queried samples, ASSFT improved performance by approximately 9.4 and 8.1 Dice points compared with RAND. Compared with the strong hybrid baseline SANN, ASSFT further achieved improvements of about 3.6 and 3.3 Dice points for $5\%$ and $10\%$ query budgets, respectively. DKD+ASD also consistently achieved the second-best performance, further confirming the effectiveness of the proposed query strategy. DKD+ASD improved Dice scores by 8.2 and 7.1 points compared with RAND for $5\%$ and $10\%$ queried samples, and achieved about 2.2–2.4 Dice points higher scores than SANN.

\subsubsection{Results on Abdominal MRI}
\begin{figure*}[!t]
\centering
\includegraphics[width=0.9\textwidth]{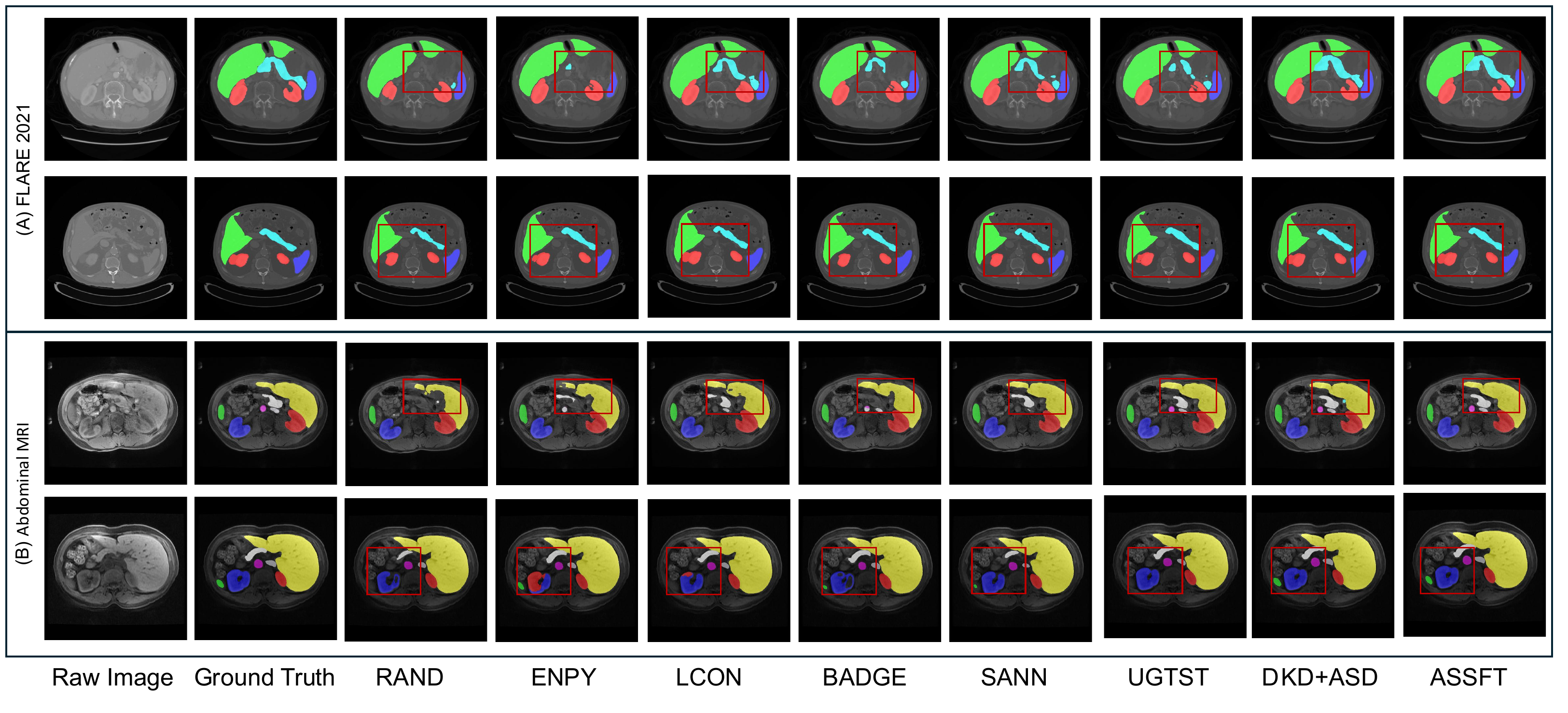}
\caption{Qualitative comparison among results of the medical vision foundation models fine-tuned by (A) $5\%$ samples from the FLARE2021 domain and (B) 3-shot from the Abdominal MRI domain queried by our methods and other SOTA methods. Red boxes mark the regions where our methods exhibit better segmentation results than SOTA methods.} 
\label{vis4}
\end{figure*}

Under the active few-shot learning setting on the Abdominal MRI dataset, ASSFT achieved the highest Dice scores across all query rounds (Table~\ref{tab5}). In the 3-shot setting, ASSFT achieved a Dice score of $83.98$, outperforming UGTST by about 5.9 points and random querying by more than 17 points. Similar improvements were observed in the 5-shot and 7-shot settings. These results demonstrated the strong capability of ASSFT for few-shot adaptation in challenging target domains. The DKD+ASD query strategy further demonstrated superior performance over SOTA query methods. In active few-shot settings, networks adapted using DKD+ASD achieved 2.5–4 Dice points higher performance than UGTST and SANN. Compared with RAND, DKD+ASD improved Dice scores by approximately 11.5–15.6 points across 3-shot, 5-shot, and 7-shot adaptation settings. Qualitative comparison provided more evidence to demonstrate the superior performance of ASSFT on cross-modality adaptation compared to other approaches (Figure~\ref{vis4}). CT-trained segmentation models adapted by three ASSFT-queried MR samples correctly segmented target anatomical structures, such as liver and kidney, while models adapted by other approaches mis-segmented these organs or misclassified them with neighboring structures.

\begin{table}[!t]
\centering
\caption{Performance comparison between our ASSFT, DKD+ASD, and other SOTA methods when adapting Med-VFMs to the Abdominal Atlas domain. The performance was evaluated using Dice scores, and the results were reported as Mean$\pm$SD. \textbf{Bold} and \underline{underline} represent the best and the second best results. Our ASSFT method achieved superior performance than other AL and ADA methods ($^*$: $p<0.01$ with the Mann-Whitney U test between ASSFT and SOTA methods).}
\label{tab4}
\resizebox{0.4\textwidth}{!}{
\begin{tabular}{c|c|c}
\toprule
\multirow{2}{*}{Methods} & \multicolumn{2}{c}{Query Budgets} \\
\cline{2-3}
           & $5\%$ & $10\%$ \\
\midrule
RAND       & 73.74$\pm$11.98 & 76.72$\pm$11.48 \\
ENPY       & 77.39$\pm$10.34 & 80.46$\pm$9.54  \\
LCON       & 74.07$\pm$10.66 & 77.88$\pm$10.90 \\
MMAR       & 75.74$\pm$11.17 & 78.33$\pm$9.81  \\
Core-set   & 79.11$\pm$10.20 & 81.44$\pm$9.61  \\
BADGE      & 79.23$\pm$10.01 & 81.48$\pm$9.52  \\
SANN       & 79.56$\pm$9.42  & 81.57$\pm$9.16  \\
UGTST      & 79.43$\pm$10.15 & 81.56$\pm$9.75  \\
CUP        & 77.36$\pm$10.46 & 79.35$\pm$9.92  \\
\midrule
DKD+ASD    & \underline{81.92}$\pm$9.31  & \underline{83.80}$\pm$8.70  \\
ASSFT      & \textbf{83.12}$^*\pm$8.16  & \textbf{84.86}$^*\pm$7.52 \\
\bottomrule
\end{tabular}}
\end{table}

\begin{table}[!t]
\centering
\caption{Performance comparison between our ASSFT, DKD+ASD, and other SOTA methods when adapting Med-VFMs to the Abdominal MRI domain. The performance was evaluated using Dice scores, and the results were reported as Mean$\pm$SD. \textbf{Bold} and \underline{underline} represent the best and the second best results. Our ASSFT method achieved superior performance than other AL and ADA methods (Lower bound $0\%$: source-training without target fine-tuning; Upper bound $100\%$: fully supervised training; $^*$: $p<0.01$ with the Mann-Whitney U test between ASSFT and SOTA methods).}
\label{tab5}
\resizebox{0.48\textwidth}{!}{
\begin{tabular}{c|c|c|c|c|c}
\toprule
\multirow{2}{*}{Methods} & Lower Bound & \multicolumn{3}{c|}{Query Budgets (Number of Iterations)} & Upper Bound\\
\cline{3-5}
         & $0\%$ & 3-shot (r=1) & 5-shot (r=2) & 7-shot (r=3) & $100\%$  \\
\midrule
RAND     & 2.01$\pm$0.39 & 66.11$\pm$15.50 & 69.77$\pm$12.71 & 76.47$\pm$11.32 & 92.95$\pm$1.14 \\
ENPY     & 2.01$\pm$0.39 & 74.10$\pm$12.77 & 79.52$\pm$11.78 & 81.05$\pm$10.12 & 92.95$\pm$1.14 \\
LCON     & 2.01$\pm$0.39 & 75.46$\pm$12.19 & 80.32$\pm$11.82 & 82.54$\pm$9.94  & 92.95$\pm$1.14 \\
MMAR     & 2.01$\pm$0.39 & 76.70$\pm$12.25 & 80.45$\pm$11.20 & 82.76$\pm$9.93  & 92.95$\pm$1.14 \\
Core-set & 2.01$\pm$0.39 & 77.94$\pm$11.15 & 81.62$\pm$10.82 & 84.21$\pm$9.45  & 92.95$\pm$1.14 \\
BADGE    & 2.01$\pm$0.39 & 77.19$\pm$11.11 & 81.70$\pm$9.47  & 84.35$\pm$9.39  & 92.95$\pm$1.14 \\
SANN     & 2.01$\pm$0.39 & 77.48$\pm$10.14 & 82.13$\pm$8.86  & 85.34$\pm$8.44  & 92.95$\pm$1.14 \\
UGTST    & 2.01$\pm$0.39 & 78.11$\pm$9.51  & 82.21$\pm$8.98  & 85.43$\pm$8.30  & 92.95$\pm$1.14 \\
CUP      & 2.01$\pm$0.39 & 74.92$\pm$12.84 & 80.22$\pm$11.92 & 81.95$\pm$10.61 & 92.95$\pm$1.14 \\
\midrule
DKD+ASD  & 2.01$\pm$0.39 & \underline{81.40}$\pm$9.41  & \underline{85.34}$\pm$8.91  & \underline{87.95}$\pm$8.16  & 92.95$\pm$1.14 \\
ASSFT    & 2.01$\pm$0.39 & \textbf{83.98}$^*\pm$8.29  & \textbf{87.63}$^*\pm$7.47  & \textbf{88.36}$^*\pm$7.32  & 92.95$\pm$1.14 \\
\bottomrule
\end{tabular}}
\end{table}

\subsection{Analysis of Generalization and Adaptation Efficiency}
The proposed ASSFT framework and DKD+ASD query strategy demonstrated strong generalization across segmentation tasks on datasets with diverse imaging modalities and varying numbers of target anatomical structures. Specifically, we observed improved segmentation performance over zero-shot segmentation under limited annotation budgets by comparing adaptation performance with lower bound performance. Furthermore, the framework and query strategy exhibit high adaptation efficiency, rapidly approaching fully supervised upper-bound performance on diverse target domains across different tasks while requiring minimal annotation effort.

\subsubsection{Results on CT Datasets}
The Med-VFM was pre-trained on CT images, and its segmentation network was subsequently adapted to target CT datasets with different anatomical structures (e.g., AMOS2022-CT and FLARE2021 datasets). 

On the FLARE2021 dataset (four-organ segmentaion), ASSFT and DKD+ASD increased the Dice score from the lower bound performance of $5.86$ to $92.52$ and $91.65$ using only $5\%$ queried samples (Table~\ref{tab3}). With $25\%$ queried samples, the adapted network reached $97.96\%$ of the fully supervised performance. 

On the AMOS2022-CT dataset (15-organ segmentation), ASSFT improved the Dice score from the lower-bound performance of $0.27$ to $80.51$ with only $5\%$ queried samples (Table~\ref{tab1}). Similarly, the DKD+ASD query strategy improved performance to $78.87$ Dice points from the lower bound performance of $0.27$. With $25\%$ queried samples, ASSFT achieved a Dice score of 87.68, corresponding to over $94\%$ of the fully supervised upper-bound performance.

\subsubsection{Results on MR Datasets}
The proposed ASSFT framework and DKD+ASD query strategy demonstrated high generalization ability on cross-modality adaptation, where Med-VFM was pre-trained on CT images and adapted to MR segmentation datasets (e.g., AMOS2022-MRI and Abdominal MRI datasets). 

On the AMOS2022-MRI dataset, ASSFT improved the Dice score by more than $51$ points from the lower bound performance of $0.46$ with only $5\%$ queried samples, and DKD+ASD query strategy improved the lower bound by over $48$ Dice points (Table~\ref{tab2}). With $30\%$ samples, ASSFT achieved over $95\%$ of the fully supervised upper-bound performance, demonstrating high adaptation efficiency. 

On the Abdominal MRI dataset, ASSFT demonstrated strong efficiency in the active few-shot learning setting (Table~\ref{tab5}). The Dice score improved from approximately $2.01$ in the lower-bound model to $83.98$ with only three labeled samples. With seven labeled samples, the adapted network achieved $88.36$ Dice, corresponding to more than $95\%$ of the upper-bound performance.

\subsection{Ablation Study on DKD and ASD Query Metrics}
To evaluate the effectiveness of the two proposed query metrics, DKD and ASD, we conducted ablation studies in which each metric was applied independently to select informative samples for network adaptation. Specifically, the top $5\%$ and $10\%$ samples ranked by each metric were selected to adapt the network on the AMOS2022-CT and AMOS2022-MRI datasets.

\subsubsection{Ablation on the Effectiveness of DKD}
To evaluate the effectiveness of DKD in identifying informative samples, we selected the top $5\%$ and $10\%$ samples with the highest (High DKD) and lowest (Low DKD) scores for adaptation to the AMOS2022-CT and AMOS2022-MRI datasets (Table \ref{tab6}). 

In the AMOS2022-CT domain, fine-tuning with high-DKD samples under both query budgets yielded improvements of approximately $11$–$12$ Dice points over low-DKD sampling and $2$–$3$ points over random selection. In the AMOS2022-MRI domain, high-DKD sampling achieved Dice scores of $45.23$ and $64.46$ at $5\%$ and $10\%$, respectively, corresponding to gains of $9$–$17$ Dice points over random sampling and $19$–$22$ points over low-DKD sampling.

These results indicate that DKD effectively captures the informativeness of target-domain samples, and that prioritizing high-DKD samples leads to substantial improvements in adaptation performance. The inferior performance of low-DKD sampling further corroborates the validity of DKD as a reliable criterion for sample selection.

We further conducted an ablation study to assess the individual contributions of PAKD and PD on the DKD metric (Table \ref{tab6}). Fine-tuning with samples selected based on PAKD or PD individually resulted in consistent improvements over random sampling, with gains of $1$–$2$ Dice points in the AMOS2022-CT domain and $6$–$15$ points in the AMOS2022-MRI domain. However, both methods remained consistently inferior to DKD, indicating that the integration of PAKD and PD provides a more comprehensive and effective measure of sample informativeness.

\begin{table}[!t]
\centering
\caption{The results of ablation study on DKD metric when adapting Med-VFMs to AMOS2022-CT and AMOS2022-MRI domains. The performance was evaluated using Dice scores, and the results were reported as Mean$\pm$SD. Adaptation by high DKD samples outperformed those by low DKD samples or randomly selected samples.}
\label{tab6}
\resizebox{0.48\textwidth}{!}{
\begin{tabular}{c|c|c|c|c}
\toprule
\multirow{2}{*}{Methods} & \multicolumn{2}{c|}{AMOS2022-CT} & \multicolumn{2}{c}{AMOS2022-MR} \\ 
\cline{2-5}
 & $5\%$ & $10\%$ & $5\%$ & $10\%$  \\
\midrule
RAND     & 73.31$\pm$15.50 & 77.57$\pm$14.65 & 35.84$\pm$27.14 & 48.06$\pm$25.24 \\
PAKD     & 74.92$\pm$14.12 & 78.74$\pm$12.98 & 42.48$\pm$21.66 & 62.55$\pm$20.21 \\
PD       & 74.80$\pm$14.25 & 78.65$\pm$13.06 & 41.91$\pm$21.90 & 62.01$\pm$20.15 \\
High DKD & 76.14$\pm$13.67 & 79.48$\pm$12.56 & 45.23$\pm$20.89 & 64.46$\pm$19.56 \\
Low DKD  & 64.48$\pm$14.72 & 68.20$\pm$14.83 & 26.86$\pm$22.67 & 42.59$\pm$20.75 \\
\bottomrule
\end{tabular}}
\end{table}

\subsubsection{Ablation on the Effectiveness of ASD}
We further evaluated the effectiveness of the ASD metric by selecting the top $5\%$ and $10\%$ samples with the highest ASD scores (High ASD) and the lowest ASD scores (Low ASD) for network adaptation in the AMOS2022-CT and AMOS2022-MRI datasets (Table~\ref{tab7}).

Across both datasets, fine-tuning with high-ASD samples consistently achieved higher Dice scores than using low-ASD samples. Specifically, in the AMOS2022-MRI dataset, adapting the network with $10\%$ high-ASD samples achieved a Dice score of $61.45$, while adapting with low-ASD samples resulted in only $39.08$. Similar improvements were observed in the AMOS2022-CT dataset. These results suggest that ASD effectively identifies samples with high segmentation difficulty that are beneficial for improving adaptation performance.

We further investigated the effect of incorporating binary masks in ASD by comparing it to a variant without foreground masking. This counterpart is equivalent to entropy-based querying (ENPY), which computes entropy across the entire image without applying temperature-scaled masking to foreground anatomical structures. Across both datasets and query budgets, high-ASD querying consistently outperformed ENPY by approximately $1$–$3$ Dice points, indicating that focusing on anatomically relevant regions improved the effectiveness of uncertainty-based sample selection.

Finally, we evaluated the effect of the dynamic temperature scaling function in ASD by comparing the results obtained using a fixed temperature ($\tau=1$) and the proposed dynamic temperature scaling function ($\tau(r)$). In both datasets, applying the dynamic temperature scaling function improved the Dice score by approximately $1$–$1.5$ points compared with ASD without temperature scaling. These results suggest that the dynamic temperature scaling mechanism further enhances the effectiveness of the ASD metric in identifying informative samples for adaptation.

\begin{table}[!t]
\centering
\caption{The results of ablation study on ASD metric when adapting Med-VFMs to AMOS2022-CT and AMOS2022-MRI domains. The performance was evaluated using Dice scores, and the results were reported as Mean$\pm$SD. Adaptation by high ASD samples outperformed those by low ASD samples or entropy-selected samples.}
\label{tab7}
\resizebox{0.48\textwidth}{!}{
\begin{tabular}{c|c|c|c|c|c}
\toprule
\multirow{2}{*}{Methods} & Temperature & \multicolumn{2}{c|}{AMOS2022-CT} & \multicolumn{2}{c}{AMOS2022-MR} \\ 
\cline{3-6}
 & function & $5\%$ & $10\%$ & $5\%$ & $10\%$  \\
\midrule
\multicolumn{2}{c|}{ASD w/o foreground masking} & 74.24$\pm$14.53 & 78.17$\pm$13.73 & 40.43$\pm$23.29 & 58.74$\pm$19.88 \\
\midrule
High ASD & \multirow{2}{*}{$\tau=1$} & 75.30$\pm$14.98 & 79.13$\pm$13.69 & 42.39$\pm$21.50 & 61.45$\pm$19.14 \\
Low ASD  & & 62.96$\pm$15.36 & 63.57$\pm$14.35 & 29.05$\pm$22.38 & 39.08$\pm$20.13 \\
\midrule
High ASD & \multirow{2}{*}{$\tau(r)$} & 75.92$\pm$14.39 & 79.41$\pm$13.24 & 44.10$\pm$20.41 & 63.77$\pm$18.48 \\
Low ASD  & & 64.73$\pm$15.44 & 66.43$\pm$13.88 & 29.64$\pm$21.35 & 40.15$\pm$19.49 \\
\bottomrule
\end{tabular}}
\end{table}

\subsection{Ablation Study on Selective Semi-supervised Fine-tuning}
We evaluated the impact of Selective Semi-supervised Fine-tuning on adaptation performance in multiple datasets. 

On the AMOS2022-CT dataset, incorporating this strategy with Active Test Time Sample Query strategy in our ASSFT method improved the performance by approximately 1-2 Dice points (Table \ref{tab1}). On the AMOS2022-MRI dataset under query budgets ranging $5\%$ to $30\%$ using the DKD+ASD, Selective Semi-supervised Fine-tuning further improved the performance by 2-3 Dice points (Table \ref{tab2}). 

On the FLARE 2021 dataset, employing Selective Semi-supervised Fine-tuning in our ASSFT method yielded improvements of approximately 0.5-1 Dice points compared to the sole use of the query strategy (Table \ref{tab3}). Finally, on Abdomen Atlas and Abdominal MRI datasets, applying Selective Semi-supervised Fine-tuning within our ASSFT method improved segmentation performance by approximately 1.2 and 2.5 Dice points, respectively (Table \ref{tab4} and \ref{tab5}). These results highlight that Selective Semi-Supervised Fine-Tuning consistently enhances adaptation performance of Med-VFMs across diverse datasets.

\subsection{Visual Assessments of DKD and ASD Metrics}
\begin{figure}
\centering
\includegraphics[width=0.48\textwidth]{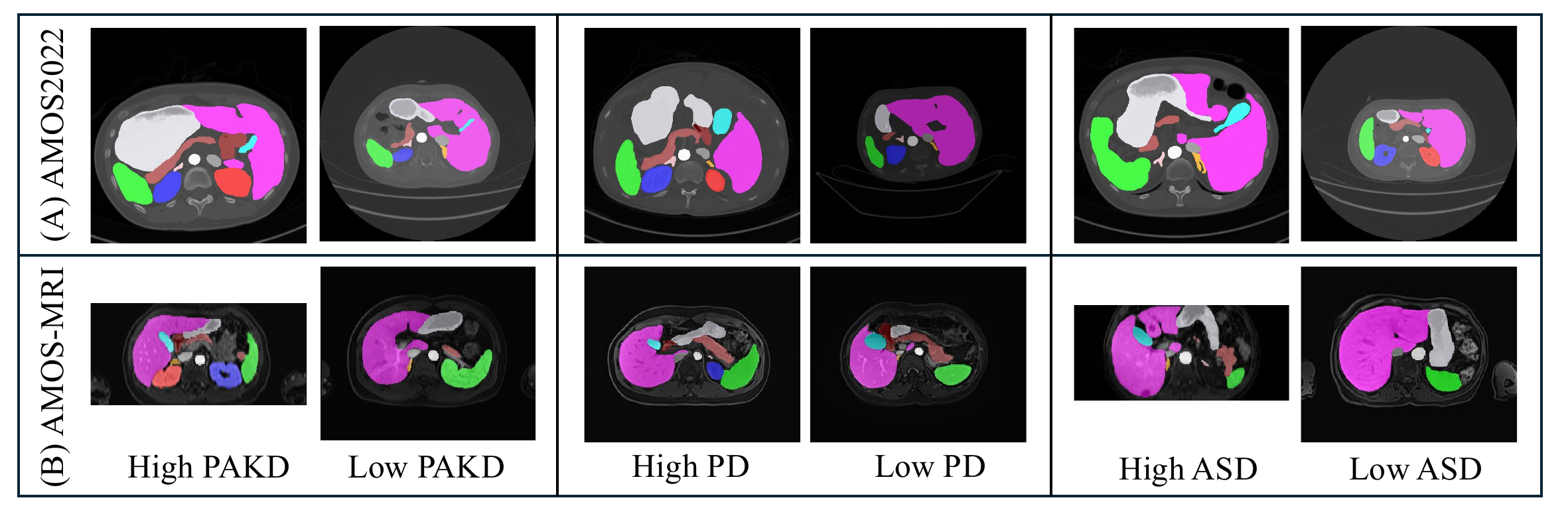}
\caption{The visualization of high- and low-PAKD samples, high- and low-PD samples, and high- and low-ASD samples in the (A) AMOS2022 and (B) AMOS-MRI datasets.} 
\label{vis5}
\end{figure}

To qualitatively illustrate which samples are deemed most informative by the three metrics, we visualized high- and low-scoring samples selected by PAKD and PD components of the DKD metric, as well as by the ASD metric (Figure \ref{vis5}). The visualizations showed that PAKD identified informative samples containing previously unlearned knowledge relative to the pre-trained model, thereby prioritizing the selection of samples with large knowledge discrepancy. In contrast, PD identified samples with substantial intra-domain diversity, promoting semantic diversity within the selected set. Additionally, ASD prioritized samples with complex anatomical structures, where the model exhibited high epistemic uncertainty, highlighting its ability to identify intrinsically challenging cases. Overall, these visualizations provide qualitative evidence that the proposed query metrics effectively identify informative samples with different characteristics for model adaptation.

\subsection{Analysis on the Effectiveness of Selective Semi-supervised Fine-tuning}
\begin{figure}
\centering
\includegraphics[width=0.48\textwidth]{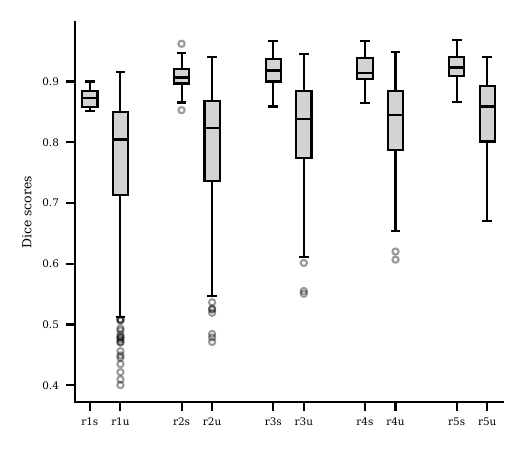}
\caption{Comparison of distributions of Dice scores from selected (s) unlabeled samples for the Selective Semi-supervised Fine-tuning and unselected (u) unlabeled samples when adapting Med-VFMs to the AMOS2022-CT domain for five rounds (r1, r2, r3, r4, and r5).} 
\label{vis6}
\end{figure}

The selection mechanism in Selective Semi-supervised Fine-tuning aims to select reliable unlabeled samples $\boldsymbol{X}_{t,u}^r$ for generating high-quality pseudo labels $\boldsymbol{Y}_{t,u}^r$ from unlabeled target sets $\mathbb{T}_u$. To demonstrate the effectiveness of this selection mechanism, we compared the distributions of Dice scores between selected unlabeled samples and unselected unlabeled samples. We adapted the Med-VFMs to the AMOS2022-CT domain for five rounds (r=1, 2, 3, 4, 5), and recorded dice scores for selected (r1s, r2s, r3s, r4s, r5s) and unselected (r1u, r2u, r3u, r4u, r5u) samples at each round (Figure \ref{vis6}). Across all rounds, selected samples consistently exhibited higher Dice scores than unselected samples.

To assess their statistical significance, we implemented Mann–Whitney U test to compare the Dice distributions of these two groups (Table \ref{tab8}). The resulting p-values were all $< 0.01$, indicating statistically significant differences at every round. These results indicate that selection mechanism in Selective Semi-supervised Fine-tuning effectively identifies reliable unlabeled samples, leading to higher-quality pseudo-labels and improved adaptation performance.

\begin{table}[!t]
\centering
\caption{The p-values of Mann-whitney u test on differences in Dice scores between selected samples and unselected samples in Selective Semi-supervised Fine-tuning when adapting Med-VFMs to the AMOS2022-CT domain.}
\label{tab8}
\resizebox{0.48\textwidth}{!}{
\begin{tabular}{c|ccccc}
\toprule
Query budgets & $5\% $ (r1) & $10\%$ (r2) & $15\%$ (r3) & $20\%$ (r4) & $25\%$ (r5) \\
\midrule
p-values      & $3.97\times10^{-6}$ & $4.53\times10^{-13}$ & $5.12\times10^{-17}$ & $5.06\times10^{-20}$ & $1.80\times10^{-22}$\\
\bottomrule
\end{tabular}}
\end{table}

\section{Discussion}
\subsection{Clinical relevance and applicability}
Med-VFMs have recently demonstrated strong performance in medical images analysis. However, the domain shifts between large-scale pre-training datasets and target clinical environments often limit their generalization, posing a major barrier to real-world deployment. Adapting Med-VFMs to new clinical domains typically requires additional annotations, which are costly and time-consuming to obtain.

The proposed ASSFT framework addresses this challenge by enabling annotation-efficient adaptation of Med-VFMs to new target domains. By actively selecting informative samples through the proposed DKD and ASD query metrics, the framework significantly reduces the annotation burden for clinical experts. Additionally, the selective semi-supervised fine-tuning mechanism leverages both labeled and unlabeled target-domain samples to further improve adaptation efficiency. These properties make ASSFT particularly suitable for real-world clinical scenarios, where only limited annotations can be obtained due to time constraints and the high cost of expert labeling.

The experimental results across five datasets demonstrate that ASSFT can rapidly improve segmentation performance from near zero-shot segmentation to near fully supervised performance with only a small number of labeled samples. This capability is particularly valuable for clinical environments where new imaging protocols, scanners, or patient populations are introduced. For example, hospitals often deploy segmentation models trained on external datasets, but their performance may degrade in new internal data due to differences in scanners or imaging protocols. In such cases, the proposed framework can efficiently adapt Med-VFMs to the local clinical data with minimal annotation effort, thereby improving the reliability and usability of automated segmentation systems.

In addition, the strong generalization of ASSFT across different imaging modalities and anatomical structures suggests its broad applicability to diverse clinical segmentation tasks. By reducing annotation requirements while maintaining high segmentation accuracy, the proposed framework facilitates the translation of Med-VFMs into practical clinical workflows for volumetric medical image analysis.

\subsection{Limitations}
Despite the promising results, several limitations of this study should be acknowledged. 

First, although the proposed framework was validated on five datasets covering multiple imaging modalities and anatomical structures, additional evaluation on a broader range of clinical datasets would further validate its generalizability and clinical applicability. In particular, future studies could investigate the performance of ASSFT on additionally modalities (e.g., Positron emission tomography).

Second, the current work focuses on volumetric medical image segmentation tasks. While the proposed active query strategy and selective semi-supervised fine-tuning framework are general in principle, their effectiveness for other medical image analysis tasks, such as detection or classification, remains to be explored.

Finally, the current ASSFT framework assumes that a small set of samples can be annotated during each active learning iteration. In real clinical workflows, however, annotation availability may vary depending on clinical resources and expert time. Moreover, weakly supervised annotations or partial organ labels may be available instead of fully supervised annotations. Future work could investigate strategies for incorporating diverse human-in-the-loop annotation systems or interactive labeling tools to better accommodate flexible annotation settings and further improve the practicality of the adaptation process. This study does not include prospective clinical validation or assessment of downstream clinical impact. While the experimental results demonstrate strong performance across multiple datasets, further evaluation in real-world clinical workflows will be necessary to establish the practical utility and clinical benefit of the proposed framework.

\section{Conclusion}
We proposed an Active Selective Semi-supervised Fine-tuning framework to efficiently adapt medical vision foundation models (Med-VFMs) to new diverse target domains for volumetric medical image segmentation. The proposed framework addresses the challenge of domain adaptation for Med-VFMs by integrating active learning with a selective semi-supervised fine-tuning mechanism. Specifically, we introduced an Active Test-Time Sample Query strategy that identifies informative target-domain samples using two complementary metrics, Diversified Knowledge Divergence and Anatomical Segmentation Difficulty, enabling effective selection of samples that maximize adaptation benefits without requiring access to source training data. Furthermore, the proposed Selective Semi-supervised Fine-tuning mechanism leverages both labeled and unlabeled target samples to further enhance adaptation performance and training efficiency. Extensive experiments across five medical image segmentation datasets demonstrate that ASSFT consistently outperforms existing active learning and active domain adaptation methods. The proposed framework not only achieves superior segmentation performance but also exhibits strong generalization across different imaging modalities and anatomical structures. Importantly, ASSFT can rapidly approach fully supervised upper-bound performance while requiring only a small number of annotated samples, highlighting its effectiveness and feasibility for annotation-efficient adaptation of Med-VFMs to new clinical domains.

\section*{Declaration of competing interests}
The authors declare that they have no known competing financial interests or personal relationships that could have appeared to influence the work reported in this paper.

\section*{Acknowledgments}
This work was supported by the National Institutes of Health grants U24 CA258483. Computations were performed using the facilities of the Washington University Research Computing and Informatics Facility (RCIF). The RCIF has received funding from NIH S10 program grants: 1S10OD025200-01A1 and 1S10OD030477-01.

\section*{Declaration of generative AI and AI-assisted technologies}
During the preparation of this work the author(s) used ChatGPT in order to improve grammer and spellings. After using this tool/service, the author(s) reviewed and edited the content as needed and take(s) full responsibility for the content of the published article.

\section*{CRediT authorship contribution statement}
JY: conceptualization, methodology, formal analysis, writing the original draft, reviewing, and editing, visualization; DM: methodology, writing, reviewing, and editing, Funding acquisition, supervision; AS: methodology, writing, reviewing, and editing, supervision.

% Use the style of numbering in square brackets.
% If nothing is used, default style will be taken.
%\begin{enumerate}[a)]
%\item 
%\item 
%\item 
%\end{enumerate}  

%\begin{itemize}
%\item 
%\item 
%\item 
%\end{itemize}  

%\begin{description}
%\item[]
%\item[] 
%\item[] 
%\end{description}  

% Uncomment and use as the case may be
%\begin{theorem} 
%\end{theorem}

% Uncomment and use as the case may be
%\begin{lemma} 
%\end{lemma}



\bibliographystyle{cas-model2-names}

\bibliography{ASFDA_VFM}

@inproceedings{settles2011theories,
  title={From theories to queries: Active learning in practice},
  author={Settles, Burr},
  booktitle={Active learning and experimental design workshop in conjunction with AISTATS 2010},
  pages={1--18},
  year={2011},
  organization={JMLR Workshop and Conference Proceedings}
}

@article{zhang2024challenges,
  title={On the challenges and perspectives of foundation models for medical image analysis},
  author={Zhang, Shaoting and Metaxas, Dimitris},
  journal={Medical image analysis},
  volume={91},
  pages={102996},
  year={2024},
  publisher={Elsevier}
}

@article{moor2023foundation,
  title={Foundation models for generalist medical artificial intelligence},
  author={Moor, Michael and Banerjee, Oishi and Abad, Zahra Shakeri Hossein and Krumholz, Harlan M and Leskovec, Jure and Topol, Eric J and Rajpurkar, Pranav},
  journal={Nature},
  volume={616},
  number={7956},
  pages={259--265},
  year={2023},
  publisher={Nature Publishing Group UK London}
}

@article{zhou2023foundation,
  title={A foundation model for generalizable disease detection from retinal images},
  author={Zhou, Yukun and Chia, Mark A and Wagner, Siegfried K and Ayhan, Murat S and Williamson, Dominic J and Struyven, Robbert R and Liu, Timing and Xu, Moucheng and Lozano, Mateo G and Woodward-Court, Peter and others},
  journal={Nature},
  volume={622},
  number={7981},
  pages={156--163},
  year={2023},
  publisher={Nature Publishing Group UK London}
}

@article{pai2025vision,
  title={Vision Foundation Models for Computed Tomography},
  author={Pai, Suraj and Hadzic, Ibrahim and Bontempi, Dennis and Bressem, Keno and Kann, Benjamin H and Fedorov, Andriy and Mak, Raymond H and Aerts, Hugo JWL},
  journal={arXiv preprint arXiv:2501.09001},
  year={2025}
}

@article{wang2026vision,
  title={Vision foundation model for 3D magnetic resonance imaging segmentation, classification, and registration},
  author={Wang, Shansong and Safari, Mojtaba and Li, Qiang and Chang, Chih-Wei and Qiu, Richard LJ and Roper, Justin and Yu, David S and Yang, Xiaofeng},
  journal={Medical Image Analysis},
  pages={103992},
  year={2026},
  publisher={Elsevier}
}

@article{zhu20263d,
  title={3D foundation model for generalizable disease detection in head computed tomography},
  author={Zhu, Weicheng and Huang, Haoxu and Tang, Huanze and Musthyala, Rushabh and Yu, Boyang and Chen, Long and Vega, Emilio and O’Donnell, Thomas and Hayek, Reya and Kuohn, Lindsey and others},
  journal={Nature Biomedical Engineering},
  pages={1--12},
  year={2026},
  publisher={Nature Publishing Group UK London}
}

@article{sun2025foundation,
  title={A foundation model for enhancing magnetic resonance images and downstream segmentation, registration and diagnostic tasks},
  author={Sun, Yue and Wang, Limei and Li, Gang and Lin, Weili and Wang, Li},
  journal={Nature Biomedical Engineering},
  volume={9},
  number={4},
  pages={521--538},
  year={2025},
  publisher={Nature Publishing Group UK London}
}

@article{tak2026generalizable,
  title={A generalizable foundation model for analysis of human brain MRI},
  author={Tak, Divyanshu and Garomsa, Biniam A and Zapaishchykova, Anna and Chaunzwa, Tafadzwa L and Climent Pardo, Juan Carlos and Ye, Zezhong and Zielke, John and Ravipati, Yashwanth and Pai, Suraj and Vajapeyam, Sri and others},
  journal={Nature Neuroscience},
  pages={1--12},
  year={2026},
  publisher={Nature Publishing Group US New York}
}

@inproceedings{siddiqui2020viewal,
  title={Viewal: Active learning with viewpoint entropy for semantic segmentation},
  author={Siddiqui, Yawar and Valentin, Julien and Nie{\ss}ner, Matthias},
  booktitle={Proceedings of the IEEE/CVF conference on computer vision and pattern recognition},
  pages={9433--9443},
  year={2020}
}

@inproceedings{wu2021redal,
  title={Redal: Region-based and diversity-aware active learning for point cloud semantic segmentation},
  author={Wu, Tsung-Han and Liu, Yueh-Cheng and Huang, Yu-Kai and Lee, Hsin-Ying and Su, Hung-Ting and Huang, Ping-Chia and Hsu, Winston H},
  booktitle={Proceedings of the IEEE/CVF international conference on computer vision},
  pages={15510--15519},
  year={2021}
}

@inproceedings{xie2022towards,
  title={Towards fewer annotations: Active learning via region impurity and prediction uncertainty for domain adaptive semantic segmentation},
  author={Xie, Binhui and Yuan, Longhui and Li, Shuang and Liu, Chi Harold and Cheng, Xinjing},
  booktitle={Proceedings of the IEEE/CVF conference on computer vision and pattern recognition},
  pages={8068--8078},
  year={2022}
}

@inproceedings{li2023heterogeneous,
  title={Heterogeneous diversity driven active learning for multi-object tracking},
  author={Li, Rui and Zhang, Baopeng and Liu, Jun and Liu, Wei and Zhao, Jian and Teng, Zhu},
  booktitle={Proceedings of the IEEE/CVF International Conference on Computer Vision},
  pages={9932--9941},
  year={2023}
}

@inproceedings{yang2024plug,
  title={Plug and play active learning for object detection},
  author={Yang, Chenhongyi and Huang, Lichao and Crowley, Elliot J},
  booktitle={Proceedings of the IEEE/CVF conference on computer vision and pattern recognition},
  pages={17784--17793},
  year={2024}
}

@inproceedings{ning2021multi,
  title={Multi-anchor active domain adaptation for semantic segmentation},
  author={Ning, Munan and Lu, Donghuan and Wei, Dong and Bian, Cheng and Yuan, Chenglang and Yu, Shuang and Ma, Kai and Zheng, Yefeng},
  booktitle={Proceedings of the IEEE/CVF international conference on computer vision},
  pages={9112--9122},
  year={2021}
}

@article{du2023diffusion,
  title={Diffusion-based probabilistic uncertainty estimation for active domain adaptation},
  author={Du, Zhekai and Li, Jingjing},
  journal={Advances in Neural Information Processing Systems},
  volume={36},
  pages={17129--17155},
  year={2023}
}

@inproceedings{zhang2024revisiting,
  title={Revisiting the domain shift and sample uncertainty in multi-source active domain transfer},
  author={Zhang, Wenqiao and Lv, Zheqi and Zhou, Hao and Liu, Jia-Wei and Li, Juncheng and Li, Mengze and Li, Yunfei and Zhang, Dongping and Zhuang, Yueting and Tang, Siliang},
  booktitle={Proceedings of the IEEE/CVF Conference on Computer Vision and Pattern Recognition},
  pages={16751--16761},
  year={2024}
}

@inproceedings{wang2023mhpl,
  title={Mhpl: Minimum happy points learning for active source free domain adaptation},
  author={Wang, Fan and Han, Zhongyi and Zhang, Zhiyan and He, Rundong and Yin, Yilong},
  booktitle={Proceedings of the IEEE/CVF Conference on Computer Vision and Pattern Recognition},
  pages={20008--20018},
  year={2023}
}

@article{li2023source,
  title={Source-free active domain adaptation via augmentation-based sample query and progressive model adaptation},
  author={Li, Shuang and Zhang, Rui and Gong, Kaixiong and Xie, Mixue and Ma, Wenxuan and Gao, Guangyu},
  journal={IEEE Transactions on Neural Networks and Learning Systems},
  year={2023},
  publisher={IEEE}
}

@inproceedings{wang2022unsupervised,
  title={Unsupervised selective labeling for more effective semi-supervised learning},
  author={Wang, Xudong and Lian, Long and Yu, Stella X},
  booktitle={European conference on computer vision},
  pages={427--445},
  year={2022},
  organization={Springer}
}

@inproceedings{he2024enhancing,
  title={Enhancing semi-supervised domain adaptation via effective target labeling},
  author={He, Jiujun and Liu, Bin and Yin, Guosheng},
  booktitle={Proceedings of the AAAI Conference on Artificial Intelligence},
  volume={38},
  number={11},
  pages={12385--12393},
  year={2024}
}

@inproceedings{kothandaraman2023salad,
  title={Salad: Source-free active label-agnostic domain adaptation for classification, segmentation and detection},
  author={Kothandaraman, Divya and Shekhar, Sumit and Sancheti, Abhilasha and Ghuhan, Manoj and Shukla, Tripti and Manocha, Dinesh},
  booktitle={Proceedings of the IEEE/CVF Winter Conference on Applications of Computer Vision},
  pages={382--391},
  year={2023}
}

@article{mahapatra2024alfredo,
  title={ALFREDO: Active Learning with FeatuRe disEntangelement and DOmain adaptation for medical image classification},
  author={Mahapatra, Dwarikanath and Tennakoon, Ruwan and George, Yasmeen and Roy, Sudipta and Bozorgtabar, Behzad and Ge, Zongyuan and Reyes, Mauricio},
  journal={Medical image analysis},
  volume={97},
  pages={103261},
  year={2024},
  publisher={Elsevier}
}

@article{wang2024dual,
  title={Dual-reference source-free active domain adaptation for nasopharyngeal carcinoma tumor segmentation across multiple hospitals},
  author={Wang, Hongqiu and Chen, Jian and Zhang, Shichen and He, Yuan and Xu, Jinfeng and Wu, Mengwan and He, Jinlan and Liao, Wenjun and Luo, Xiangde},
  journal={IEEE Transactions on Medical Imaging},
  year={2024},
  publisher={IEEE}
}

@inproceedings{luo2024uncertainty,
  title={An uncertainty-guided tiered self-training framework for active source-free domain adaptation in prostate segmentation},
  author={Luo, Zihao and Luo, Xiangde and Gao, Zijun and Wang, Guotai},
  booktitle={International Conference on Medical Image Computing and Computer-Assisted Intervention},
  pages={107--117},
  year={2024},
  organization={Springer}
}

@inproceedings{wang2024advancing,
  title={Advancing uwf-slo vessel segmentation with source-free active domain adaptation and a novel multi-center dataset},
  author={Wang, Hongqiu and Luo, Xiangde and Chen, Wu and Tang, Qingqing and Xin, Mei and Wang, Qiong and Zhu, Lei},
  booktitle={International Conference on Medical Image Computing and Computer-Assisted Intervention},
  pages={75--85},
  year={2024},
  organization={Springer}
}

@inproceedings{yang2025active,
  title={Active Source-Free Cross-Domain and Cross-Modality Adaptation for Volumetric Medical Image Segmentation by Image Sensitivity and Organ Heterogeneity Sampling},
  author={Yang, Jin and Yu, Xiaobing and Qiu, Peijie and Marcus, Daniel and Sotiras, Aristeidis},
  booktitle={International Conference on Medical Image Computing and Computer-Assisted Intervention},
  pages={3--12},
  year={2025},
  organization={Springer}
}

@article{hiasa2019automated,
  title={Automated muscle segmentation from clinical CT using Bayesian U-Net for personalized musculoskeletal modeling},
  author={Hiasa, Yuta and Otake, Yoshito and Takao, Masaki and Ogawa, Takeshi and Sugano, Nobuhiko and Sato, Yoshinobu},
  journal={IEEE transactions on medical imaging},
  volume={39},
  number={4},
  pages={1030--1040},
  year={2019},
  publisher={IEEE}
}

@article{nath2020diminishing,
  title={Diminishing uncertainty within the training pool: Active learning for medical image segmentation},
  author={Nath, Vishwesh and Yang, Dong and Landman, Bennett A and Xu, Daguang and Roth, Holger R},
  journal={IEEE Transactions on Medical Imaging},
  volume={40},
  number={10},
  pages={2534--2547},
  year={2020},
  publisher={IEEE}
}

@article{zhao2021dsal,
  title={Dsal: Deeply supervised active learning from strong and weak labelers for biomedical image segmentation},
  author={Zhao, Ziyuan and Zeng, Zeng and Xu, Kaixin and Chen, Cen and Guan, Cuntai},
  journal={IEEE journal of biomedical and health informatics},
  volume={25},
  number={10},
  pages={3744--3751},
  year={2021},
  publisher={IEEE}
}

@article{gaillochet2023active,
  title={Active learning for medical image segmentation with stochastic batches},
  author={Gaillochet, M{\'e}lanie and Desrosiers, Christian and Lombaert, Herv{\'e}},
  journal={Medical Image Analysis},
  volume={90},
  pages={102958},
  year={2023},
  publisher={Elsevier}
}

@article{li2023hal,
  title={HAL-IA: A hybrid active learning framework using interactive annotation for medical image segmentation},
  author={Li, Xiaokang and Xia, Menghua and Jiao, Jing and Zhou, Shichong and Chang, Cai and Wang, Yuanyuan and Guo, Yi},
  journal={Medical Image Analysis},
  volume={88},
  pages={102862},
  year={2023},
  publisher={Elsevier}
}

@inproceedings{zhou2024sbc,
  title={SBC-AL: Structure and Boundary Consistency-Based Active Learning for Medical Image Segmentation},
  author={Zhou, Taimin and Yang, Jin and Cui, Lingguo and Zhang, Nan and Chai, Senchun},
  booktitle={International Conference on Medical Image Computing and Computer-Assisted Intervention},
  pages={283--293},
  year={2024},
  organization={Springer}
}

@article{shu2025active,
  title={An active learning model based on image similarity for skin lesion segmentation},
  author={Shu, Xiu and Li, Zhihui and Tian, Chunwei and Chang, Xiaojun and Yuan, Di},
  journal={Neurocomputing},
  pages={129690},
  year={2025},
  publisher={Elsevier}
}

@inproceedings{he2015delving,
  title={Delving deep into rectifiers: Surpassing human-level performance on imagenet classification},
  author={He, Kaiming and Zhang, Xiangyu and Ren, Shaoqing and Sun, Jian},
  booktitle={Proceedings of the IEEE international conference on computer vision},
  pages={1026--1034},
  year={2015}
}

@article{ji2022amos,
  title={Amos: A large-scale abdominal multi-organ benchmark for versatile medical image segmentation},
  author={Ji, Yuanfeng and Bai, Haotian and Ge, Chongjian and Yang, Jie and Zhu, Ye and Zhang, Ruimao and Li, Zhen and Zhanng, Lingyan and Ma, Wanling and Wan, Xiang and others},
  journal={Advances in neural information processing systems},
  volume={35},
  pages={36722--36732},
  year={2022}
}

@article{ma2022fast,
  title={Fast and low-GPU-memory abdomen CT organ segmentation: the flare challenge},
  author={Ma, Jun and Zhang, Yao and Gu, Song and An, Xingle and Wang, Zhihe and Ge, Cheng and Wang, Congcong and Zhang, Fan and Wang, Yu and Xu, Yinan and others},
  journal={Medical Image Analysis},
  volume={82},
  pages={102616},
  year={2022},
  publisher={Elsevier}
}

@article{li2025well,
  title={How well do supervised 3d models transfer to medical imaging tasks?},
  author={Li, Wenxuan and Yuille, Alan and Zhou, Zongwei},
  journal={arXiv preprint arXiv:2501.11253},
  year={2025}
}

@article{zhou2025mrannotator,
  title={MRAnnotator: multi-anatomy and many-sequence MRI segmentation of 44 structures},
  author={Zhou, Alexander and Liu, Zelong and Tieu, Andrew and Patel, Nikhil and Sun, Sean and Yang, Anthony and Choi, Peter and Lee, Hao-Chih and Tordjman, Mickael and Deyer, Louisa and others},
  journal={Radiology Advances},
  volume={2},
  number={1},
  pages={umae035},
  year={2025},
  publisher={Oxford University Press}
}

@article{li2006confidence,
  title={Confidence-based active learning},
  author={Li, Mingkun and Sethi, Ishwar K},
  journal={IEEE transactions on pattern analysis and machine intelligence},
  volume={28},
  number={8},
  pages={1251--1261},
  year={2006},
  publisher={IEEE}
}

@inproceedings{wang2014new,
  title={A new active labeling method for deep learning},
  author={Wang, Dan and Shang, Yi},
  booktitle={2014 International joint conference on neural networks (IJCNN)},
  pages={112--119},
  year={2014},
  organization={IEEE}
}

@inproceedings{yang2017suggestive,
  title={Suggestive annotation: A deep active learning framework for biomedical image segmentation},
  author={Yang, Lin and Zhang, Yizhe and Chen, Jianxu and Zhang, Siyuan and Chen, Danny Z},
  booktitle={Medical Image Computing and Computer Assisted Intervention- MICCAI 2017: 20th International Conference, Quebec City, QC, Canada, September 11-13, 2017, Proceedings, Part III 20},
  pages={399--407},
  year={2017},
  organization={Springer}
}

@inproceedings{sener2018active,
title={Active Learning for Convolutional Neural Networks: A Core-Set Approach},
author={Ozan Sener and Silvio Savarese},
booktitle={International Conference on Learning Representations},
year={2018},
url={https://openreview.net/forum?id=H1aIuk-RW},
}

@inproceedings{Ash2020Deep,
title={Deep Batch Active Learning by Diverse, Uncertain Gradient Lower Bounds},
author={Jordan T. Ash and Chicheng Zhang and Akshay Krishnamurthy and John Langford and Alekh Agarwal},
booktitle={International Conference on Learning Representations},
year={2020}
}

@article{chen2025active,
  title={Active learning for cross-modal cardiac segmentation with sparse annotation},
  author={Chen, Zihang and Zhao, Weijie and Liu, Jingyang and Xie, Puguang and Hou, Siyu and Nian, Yongjian and Yang, Xiaochao and Ma, Ruiyan and Ding, Haiyan and Xiao, Jingjing},
  journal={Pattern Recognition},
  volume={162},
  pages={111403},
  year={2025},
  publisher={Elsevier}
}

@article{wang2024patch,
  title={A patch distribution-based active learning method for multiple instance Alzheimer's disease diagnosis},
  author={Wang, Tianxiang and Dai, Qun},
  journal={Pattern Recognition},
  volume={150},
  pages={110341},
  year={2024},
  publisher={Elsevier}
}

@article{deng2025brain,
  title={Brain foundation models with hypergraph dynamic adapter for brain disease analysis},
  author={Deng, Zhongying and Wang, Haoyu and Huang, Ziyan and Zhang, Lipei and Aviles-Rivero, Angelica I and Liu, Chaoyu and He, Junjun and Kourtzi, Zoe and Sch{\"o}nlieb, Carola-Bibiane},
  journal={Pattern Recognition},
  pages={112595},
  year={2025},
  publisher={Elsevier}
}



\end{document}